# Using potential derivatives to reduce the cost of rectangular collocation: a general and versatile approach for computational vibrational spectroscopy without a potential energy surface

Moksh Bhateja,[1,a] Tucker Carrington Jr,[1,b] and Sergei Manzhos [2,c]

[1] Department of Chemistry 90 Bader Lane Queen's University, Kingston, Ontario K7L 3N6, Canada

[2] School of Materials and Chemical Technology, Institute of Science Tokyo, Ookayama 2-12-1, Meguro-ku, Tokyo 152-8552 Japan

## Abstract

We modify the exact Cartesian kinetic energy operator, rectangular collocation approach of Manzhos and Carrington [*J. Chem. Phys.* **145** (2016) 224110] by using information contained in potential derivatives to decrease the number of required potential points. We show that this makes it possible to substantially reduce the number of collocation points, which is important, in particular, when energies are computed directly ab initio. Vibrational band intensities are also computed. We first demonstrate the method on $Cl_2CO$ and compare to high-accuracy Smolyak collocation calculations on an accurate potential energy surface (PES). Sub-cm$^{-1}$ accuracy is obtained with a few thousand Gaussian basis functions and potential points. A spectrum computed directly from DFT energies and a comparison to GVPT2 (generalized vibrational perturbation theory) are also presented. We then consider $LiCO_2$ as a more anharmonic case, for which a quartic force field potential is less reliable, and which also serves as a surrogate model for a catalytic system (mimicking $CO_2$ bond weakening by charge donation). The rectangular collocation method is used to compute the spectrum of $LiCO_2$ directly from ab initio energies and to compare to GVPT2. The method is implemented as a Claude Skill and is easily usable by anyone on any molecule. The method is a general and versatile tool for anharmonic computational vibrational spectroscopy. It does not require a PES and no coupling is neglected.

[a] Current address: Max Planck Institute for the Physics of Complex Systems (MPI-PKS), 01187 Dresden, Germany.

[b] Email: tucker.carrington@queensu.ca

[c] E-mail: Manzhos.s.aa@m.titech.ac.jp

# 1 Introduction

Computing vibrational spectra is an important goal in chemical physics. It is important for interpreting and understanding gas phase spectroscopic experiments and also used for species identification in various applications. Vibrational spectroscopies are among the most widely used characterization techniques in all states of matter.[1–4] Computing an accurate vibrational spectrum with an established variational method is difficult for three reasons. 1) One requires a potential energy surface (PES). To obtain a PES, it is necessary to calculate ab initio energies at a large number of points and *then* to fit a function to the points. 2) The kinetic energy operator (KEO) is complicated and/or not known in the best coordinates. 3) It is costly to compute the spectrum, both in terms of memory and in terms of CPU time. Both computing matrix elements and solving the variational eigenvalues problem are expensive. The method we present in this paper resolves all three of these impediments to computing spectra.

Owing to the three impediments, it is common to resort to approximate methods. The simplest approach neglects all anharmonicity and coupling. However, anharmonicity and coupling can shift vibrational transitions by dozens or hundreds of $cm^{-1}$, significantly affecting assignment.[1,4–10] To account for these effects, one needs a PES. It is often approximated with a Taylor series (force field)[11] or a high-dimensional model representation (HDRM)[12] (also called an order-of-coupling approximation[13]) or an n-mode representation.[14] Quartic force fields are often used with VPT2 ($2^{nd}$ order vibrational perturbation theory).[15–19] An n-mode representation is often used with VSCF (vibrational self-consistent field)[14,20–24] and VCI (vibrational configuration interaction).[14,24–26]

Despite the impediments, it *is* possible to compute a numerically exact spectrum for a molecule with fewer than about 7 atoms.[7,27–29] The impediments, however, make the calculation difficult. A key problem is that a variational method is used and Hamiltonian matrix elements are evaluated with quadrature. A lot is known about how to reduce the required number of basis functions (size of the variational Hamiltonian matrix), but the quadrature problem remains debilitating.[27] Tensor product quadratures are common and for molecules with more than three atoms there are millions of quadrature points. Accurate quadrature schemes with fewer points do exist,[30,31] but the problem remains serious. When using general multi-dimensional basis functions, it is not known how to find an accurate quadrature with about as many points as the basis set. Because the number of quadrature points is large, one almost always fits a PES and then evaluates it at the quadrature points. It would be infeasible to compute ab initio energies at all of the quadrature points.[7,27,32–34] In many applications in which vibrational spectroscopy is important, PESs are not available. This holds, for example, for the vast majority of materials studied experimentally. A case in point is adsorbed molecules (e.g. on catalytic surfaces) where only for a small fraction of the molecule-surface combinations studied experimentally do there exist PESs. In these applications, experimental resolution is typically several $cm^{-1}$.[3,35–40] Even with this limited accuracy, anharmonicity and coupling effects are important because they are on the order of $10^1$-$10^2$ $cm^{-1}$ and can affect species assignment.[9,10,37,41–44] In most materials science applications, computational methods used for vibrational spectra assignment either use a harmonic PES (computed with DFT (density functional theory)) or VPT2, VSCF, or VCI.[14,42,44–48] In many applications, it is just too complicated and too costly to do better calculations using current methods. There is a clear need in materials science applications for a simple method with which one can compute spectra, accounting for anharmonicity and coupling, without a PES, with as few approximations as possible. We provide such a method in this paper.

Instead of a variational method, we use a collocation method.[49,50] The Schrödinger equation is solved by expanding wavefunctions in a basis and demanding that it be satisfied at a set of collocation points. There are no integrals and hence quadrature is not required. Another advantage of collocation is that to make the matrix for the KEO one needs only to act with the KEO on basis functions and then evaluate the resulting functions at points. This makes it possible to use a KEO in space-fixed (SF) Cartesian coordinates, also when basis functions are functions of internal coordinates. Owing to the fact that one needs only to evaluate the function obtained by applying the KEO to basis functions at points, one is free to use a different space-fixed frame for every collocation point. In a previous paper.[51] we called these space-fixed frames rotated space-fixed frames (RSF). By using space-fixed frames, we obviate both the need to derive a KEO and the need to deal with complicated KEOs. The space-fixed KEO is exact.

The most obvious collocation method uses as many points as basis functions.[52] Accuracy can be improved by using more points than basis functions.[49,53] There are two ways to do this. The most obvious uses the standard pseudoinverse.[54] The other option is to chop the inverse of a larger $\boldsymbol{B}$ matrix (see Eq. 9 below).[55] Both methods are called rectangular collocation (RC). Regardless of which method is used, the size of the collocation point set is related to (and not much bigger than) the number of basis functions. When it is possible to find a small basis with which one can accurately represent wavefunctions, only a small number of collocation points will be required. In this paper, we use a distributed Gaussian basis (DGB),[56–58] but other bases could be used with the space-fixed RC method. We show that with a small DGB, and therefore a small point set, it is possible to compute accurate spectra. Importantly, the point set is small enough that it is possible to compute ab initio energies at the points, obviating the need for a PES.[9,10,41,59–61] We show that the DGB is small enough that the matrix eigenvalue problem can be easily solved. Although in this paper we use a DGB, the method we propose is general and can be used with any collocation point distribution[62–64] and with any type of basis functions, including difficult-to-integrate or non-integrable functions.[64,65] Using more points than basis functions facilitates basis optimization,[9,41,59–61,66] which can be used to reduce the required number of basis functions and collocation points. For large systems, a subset of the coordinates can be used, if it is large enough to compute the properties of interest [41,60]. To date, calculations with the method have been shown to have errors on the order of 1 $cm^{-1}$, which is sufficient for most materials science applications.

Standard (square) collocation methods solve the Schrödinger equation by constraining wavefunctions (in a given basis) to be exact at a set of points. In this work, we improve the SF rectangular collocation method by imposing additional constraints obtained by using potential derivatives. This substantially reduces the required number of collocation points / ab initio calculations which is especially important when energies at points are computed ab initio and not from a PES, e.g. in materials science applications. The fact that it is cheap to obtain not only energies, but also potential gradients, from many ab initio programs and for several quantum chemistry methods, makes this is a useful strategy. We also calculate vibrational band intensities.

When using RC, one needs to optimize a number of calculation parameters including parameters that determine the collocation point distribution, DGB function placement, and other basis parameters. While this might appear to be an additional complication, in practice it is not, as the method is naturally amenable to automated optimization which is facilitated by the rectangular nature of the collocation equation.[9,41,59–61,66] This is an important advantage in the age of AI: we implement the method as a Claude Skill that greatly facilitates its use by non-experts, in particular, enabling automatic coordinate conversions and the tuning of various parameters (including numbers of basis functions and collocation points, basis function parameters, and configuration space sampling parameters) for convergence using, in particular, the residual of the rectangular collocation equation. The method is a general and versatile tool for computing accurate vibrational spectra without a potential energy surface.

We demonstrate the accuracy and efficiency of the method by doing calculations on $Cl_2CO$ and $LiCO_2$. $Cl_2CO$ is chosen because we can compare to high-accuracy Smolyak collocation calculations on an accurate PES.[34] Sub-cm$^{-1}$ accuracy is obtained with only a few thousand basis functions and points. A $Cl_2CO$ spectrum is also computed *directly* from DFT energies and gradients. Both spectra are compared to GVPT2 (generalized vibrational perturbation theory) spectra. Calculations for $LiCO_2$ are done directly from DFT energies and gradients; this example showcases the utility of the method for a system similar to those of interest in materials science. It is a more anharmonic case, for which a quartic PES is less reliable, and which also serves as a surrogate model for a catalytic system (mimicking $CO_2$ bond weakening by charge donation).

# 2 Methods

## *2.1 Rectangular collocation with exact Cartesian kinetic energy operator and potential derivatives*

The vibrational Schrödinger equation

$$\left(\hat{T}(\boldsymbol{q}) + V(\boldsymbol{q})\right)\psi_i(\boldsymbol{q}) = E_i\psi_i(\boldsymbol{q}) \tag{1}$$

where $\hat{T}(\boldsymbol{q})$ is the kinetic energy operator (KEO), $V(\boldsymbol{q})$ is the potential energy operator, and $E_i$ is the energy, is solved by expanding the wavefunctions in $N$ basis functions that are anisotropic Gaussian functions $\phi_j(\boldsymbol{q}) = \exp\left[-\frac{1}{2}\sum_k \frac{\left(q_k - q_k^{(j)}\right)^2}{\beta_k^2}\right]$ in dimensionless normal coordinates $\boldsymbol{q}$,

$$\psi_i(\boldsymbol{q}) = \sum_{j=1}^{N} c_{ji}\ \phi_j(\boldsymbol{q}). \tag{2}$$

The coefficients are obtained by “solving” a rectangular eigenvalue problem,

$$\sum_j [\hat{T}\phi_j(\boldsymbol{q}^{(m)})]c_{ji} + \sum_j V(\boldsymbol{q}^{(m)})\phi_j(\boldsymbol{q}^{(m)})c_{ji} = \sum_j \phi_j(\boldsymbol{q}^{(m)})\, c_{ji}\, E_{ii}, \qquad m = 1, \dots, M. \tag{3}$$

$M > N$ and $\boldsymbol{q}^{(m)}$ are collocation points. Gaussian basis functions, with variational[56,58,67] and with collocation[51,62,68] methods, have been used to compute spectra for many years.

The *M* collocation points are the first *M* points, for which the energy is less than $V_{max}$, of an $R_d$ ($d$ = 6) pseudorandom sequence of points in a *D*-dimensional cube. The sides of the cube start at the negative value of $q_k$ at which the potential is $V_{max}$ and end at the positive value of $q_k$ at which the potential is $V_{max}$. An $R_6$ pseudorandom sequence (a type of Kronecker sequence [69]) is used because it is uniform in the cube. It has less clustering than the Sobol sequence[70] used in previous calculations.[53,62,64] Flynn and Mandelshtam used quasi-random points and discussed the clustering problem.[68] In this paper, we do not use an accept/reject method to obtain a distribution with more points near the equilibrium geometry. When tuning $V_{max}$, there is no need to regenerate the pseudorandom sequence for each trial $V_{max}$, one simply generates the sequence with a sufficiently high $V_{max}$ once. The *N* Gaussian centers are chosen from a separate segment of the $R_6$ sequence (starting from index 20,000,000) and points with energy less than $V'_{max} < V_{max}$ are used.

We use an exact $J = 0$ kinetic energy operator. At different points we use different space-fixed frames, but at all points the KEO is simply the Cartesian Laplacian

$$\hat{T} = -\frac{1}{2}\sum_{i=1}^{3N_{at}} \frac{1}{m_i}\frac{\partial^2}{\partial x_i^2} \tag{4}$$

where $m_i$ are atomic masses. We must act with this KEO on basis functions that depend on the normal coordinates. This is done using central finite differences in Cartesian coordinates. A five-point stencil is typically sufficient and is used in this work, with a finite difference step size of $10^{-3}$ Bohr (results are not sensitive to this parameter). To do this, we must find the normal coordinates that correspond to Cartesian points $\boldsymbol{x}$, i.e. $\boldsymbol{q}(\boldsymbol{x})$. The space-fixed frame we use at each point has the same orientation as the Eckart frame (for the equilibrium configuration) at that point. We apply the KEO to $\phi(\boldsymbol{q}(\boldsymbol{x}))$.

If derivatives of the potential are easily available (which is the case for many ab initio methods and naturally for analytic PESs), one can use the derivative information to increase the number of equations without increasing *N*. This is done by differentiating Eq. (3). If $\psi_i(\boldsymbol{q})$ is an exact eigenfunction, the residual $R_i = (\hat{T} + V - E_i)\psi_i$ vanishes identically, and so does its gradient. Acting with $\partial/\partial q_k$ on both sides of Eq. 1 and then evaluating at a point, one obtains

$$\sum_j [\hat{T}\chi_{kj}(\boldsymbol{q}^{(m)})]c_{ji} + \sum_j V(\boldsymbol{q}^{(m)})\chi_{kj}(\boldsymbol{q}^{(m)})c_{ji} + \sum_j V_k(\boldsymbol{q}^{(m)})\phi_j(\boldsymbol{q}^{(m)})c_{ji} = \sum_j \chi_{kj}(\boldsymbol{q}^{(m)})\, c_{ji}\, E_{ii}, \qquad k = 1, \dots, n_q, \tag{6}$$

with

$$\chi_{kj} = \frac{\partial \phi_j}{\partial q_k}, \qquad V_k = \frac{\partial V}{\partial q_k}. \tag{7}$$

Because $\hat{T}$ commutes with $\partial/\partial q_k$, the kinetic term is evaluated as

$$\hat{T}\chi_{kj} = \frac{\partial}{\partial q_k}(\hat{T}\phi_j). \tag{8}$$

One might choose to differentiate with respect to only a subset of the coordinates and therefore $n_q$ in Eq. 6 might be less than *D*. The derivative in Eq. 8 is evaluated by finite difference (five-point stencil). When computing $\hat{T}\phi_j$, the direction of the rotated space-fixed frame does not change when Cartesian coordinates are increased or decreased during the finite difference procedure. Only $V$ and $\nabla V$ at the collocation points are needed.

To build the matrices of the matrix eigenvalue problem we solve, we stack together rows obtained from the Schrödinger equation

$$B_{mj} = \phi_j(\boldsymbol{q}^{(m)}), \qquad A_{mj} = [(\hat{T} + V)\phi_j](\boldsymbol{q}^{(m)}), \tag{9}$$

and *D* rows made with potential derivatives,

$$B_{mj}^{(k)} = \chi_{kj}(\boldsymbol{q}^{(m)}), \qquad A_{mj}^{(k)} = [(\hat{T} + V)\chi_{kj}](\boldsymbol{q}^{(m)}) + V_k(\boldsymbol{q}^{(m)})\,\phi_j(\boldsymbol{q}^{(m)}). \tag{10}$$

The (rectangular) eigenvalue problem is $\tilde{A}\,\mathbf{c} = E\,\tilde{B}\,\mathbf{c}$. It is best to modify the matrices $\tilde{A}$ and $\tilde{B}$. The influence of the *D* rows made with potential derivatives shifts up the eigenvalues because for points far from equilibrium the potential derivatives are large. To mitigate this problem, we weight the *D* rows made with potential derivatives using

$$\lambda_m = \left(1 - \frac{V(X_m)}{V_{\max}}\right)^p, \tag{11}$$

with $p \approx 1 - 3$ being optimal for the molecules considered here. The weight is smaller for points at higher energy. Setting $\lambda_m = 0$ at selected points uses gradients only at those points.

The rectangular generalized eigenproblem, with derivative rows weighted by $\sqrt{\lambda_m}$, has $M(1 + D)$ rows and the modified $\tilde{A}$ and $\tilde{B}$ matrices, is:

$$\tilde{A}\,\mathbf{c} = E\,\tilde{B}\,\mathbf{c}, \qquad \tilde{B} = \begin{pmatrix} B \\ \Lambda^{1/2}B^{(1)} \\ \vdots \\ \Lambda^{1/2}B^{(n_q)} \end{pmatrix}, \qquad \tilde{A} = \begin{pmatrix} A \\ \Lambda^{1/2}A^{(1)} \\ \vdots \\ \Lambda^{1/2}A^{(n_q)} \end{pmatrix}. \tag{12}$$

To solve the rectangular generalized eigenproblem, we left multiply by $\tilde{B}^T$ to obtain a square eigenvalue problem with $N \times N$ matrices. The square matrices are computed, without storing the large rectangular matrices, by accumulating their elements point by point to obtain,

$$S = \sum_m \left[\mathbf{b}_m^T \mathbf{b}_m + \lambda_m \sum_k \mathbf{b}_m^{(k)T}\,\mathbf{b}_m^{(k)}\right], \qquad H = \sum_m \left[\mathbf{b}_m^T \mathbf{a}_m + \lambda_m \sum_k \mathbf{b}_m^{(k)T}\,\mathbf{a}_m^{(k)}\right]. \tag{13}$$

The final square generalized matrix equation is

$$\boldsymbol{H}\,\mathbf{c} = \boldsymbol{E}\,\boldsymbol{S}\,\mathbf{c}, \tag{14}$$

where $\boldsymbol{H}$ is in general a nonsymmetric. The lowest $n_{ev}$ eigenpairs (corresponding to the desired number of vibrational levels) are obtained using ARPACK in shift-invert mode, i.e. we compute eigenvalues of $(H - \sigma S)^{-1}S$:

$$(H - \sigma S)^{-1} S\, \mathbf{c} = \nu\, \mathbf{c}, \qquad E = \sigma + \frac{1}{\nu}.$$

(15)

$\sigma$ is just below the zero-point energy and the linear system is solved by doing an LU decomposition. The condition number of $\boldsymbol{S}$ is large. Consequently, the Lebesgue constant, which enters into the error bound for collocation energies is also large.[71] However, because the bound is small when the basis error is small, the collocation energies are expected to be accurate.

*2.2 Vibrational band intensities*

From the collocation calculation, we obtain wavefunctions as linear combinations of basis functions that are functions of the normal coordinates. To compute vibrational band intensities, we need to calculate

$$\boldsymbol{\mu}_{0f} = \int \psi_0(\boldsymbol{q})\, [\boldsymbol{\mu}(\boldsymbol{q})]\, \psi_f(\boldsymbol{q})\, \rho(\boldsymbol{q})\, d\boldsymbol{q},$$

(16)

where $\psi_0(\boldsymbol{q})$ is the ground state and $\psi_f$ is some final state. The wavefunctions depend on the normal coordinates so the integral is an integral over $\mathbf{q}$. The appropriate volume element is determined from the Jacobian for the transformation between the $3N_{at}$ Cartesian coordinates and the set of $3N_{at}$ coordinates composed of the $3N_{at} - 6$ normal coordinates, 3 translational coordinates, and 3 rotational coordinates. The vibrational part of this Jacobian is[72]

$$\rho(\boldsymbol{q}) = \left| \det \left[ \frac{\partial \boldsymbol{X}}{\partial t_a}, \mathbf{e}_a \times \boldsymbol{X}, \frac{\partial \boldsymbol{X}}{\partial q_k} \right] \right|,$$

(17)

where $t_a$ are translational coordinates. In the middle, $\mathbf{e}_a \times \boldsymbol{X}$ is shorthand for a stack of $N_{at}$ 3×3 matrices $\mathbf{e}_a \times \boldsymbol{X}_i$ and is the rotational part of the Jacobian.

To improve the accuracy of the intensities, the imaginary part of the collocation wavefunctions $\psi_f = \sum_j c_{fj}\, \phi_j$ is dropped and they are Löwdin-orthonormalized, using the $\rho(\boldsymbol{q})$ metric. Finally, dipole matrix elements and vibrational band intensities are computed from

$$\boldsymbol{\mu}_{0f} = \int \psi_0(\boldsymbol{q})\, [\boldsymbol{\mu}(\boldsymbol{q}) - \bar{\boldsymbol{\mu}}_0]\, \psi_f(\boldsymbol{q})\, \rho(\boldsymbol{q})\, dq,$$

$$\bar{\boldsymbol{\mu}}_0 = \int \psi_0^2(\boldsymbol{q})\, \boldsymbol{\mu}(\boldsymbol{q})\, \rho(\boldsymbol{q})\, d\boldsymbol{q},$$
$$I_{0f} = \frac{8\pi^3 N_A}{3hc\,(4\pi\varepsilon_0)}\, \tilde{\nu}_{0f} \left|\boldsymbol{\mu}_{0f}\right|^2,$$

(18)

where $\tilde{\nu}_{0f}$ is the transition wavenumber. We are assuming that the energies of excited states are much higher than the ground state energy, relative to $kT$, in this case the intensity is independent of temperature. Subtracting $\bar{\boldsymbol{\mu}}_0$ removes residual error due to non-orthogonality. The integrals are evaluated with a quasi-Monte-Carlo quadrature for which the first 400,000 points with energy less than $V_{max}$ are chosen from the $R_6$ sequence. In this paper, when we use a PES, we use a linear dipole approximation

$$\boldsymbol{\mu}(q) \approx \boldsymbol{\mu}_0 + \sum_k \left(\frac{\partial \boldsymbol{\mu}}{\partial q_k}\right)_0 q_k,$$

(19)

and ab initio dipole derivatives. When computing a spectrum directly from ab initio energies, we fit dipole moment surfaces for the three dipole components, using dipole component values computed at the $M$ collocation points.

## *2.3 Potential energy calculations and method implementation*

A key advantage of the method is that the number or collocation points is small enough that ab initio energies can be computed at those points rather than first fitting a PES. However, the method can of course also be used with a PES. We report results from tests with a PES and tests with ab initio energies and gradients computed at collocation points. The test with a PES is for $Cl_2CO$. It is chosen because accurate energies, with which we can compare, are known.[34] We also compute spectra of both $Cl_2CO$ and $LiCO_2$ (for which no accurate PES is available) directly from ab initio energies and gradients. We use DFT[73,74] with the B3LYP[75] functional and aug-cc-pVDZ and -pVTZ basis sets (UB3LYP in the case of $LiCO_2$). As the purpose is only to demonstrate the efficacy of the vibrational Schrödinger solver, the accuracy of the ab initio method is not critical. DFT calculations were done in PySCF.[76,77] For phosgene, dipole moment values were also calculated at the CCSD(T)[78,79] level of theory, also in PySCF.

The method was taught to the Claude AI program (using the Claude Opus 5.5 large language model (LLM)[80]) and is available to users as a Claude Skill (https://github.com/sergeimanzhos/rectcolloc-vib ) that can be applied to any molecule. Using natural language instructions, users can request that Claude use the Skill with ab initio codes accessible to the user with any levels of theory, or a user-supplied PES function. Importantly, Claude optimizes computational parameters: basis function widths for each coordinate, $V_{max}$ and $V'_{max}$, as well as the *M/N* ratio. This is done using the norm of the residual evaluated at a set of validation points. The validation points are drawn from the same $R_6$ sequence, also with energy less than $V_{max}$. To make sure that they do not overlap with collocation points for any *M*, they are drawn from a continuous portion of the sequence starting from a high index (much higher than plausible *M* values), in this case 10,000,000. For energy level $i$, the residual is[66]

$$r_i = \frac{\|(\hat{H} - E_i)\psi_i\|}{\|\psi_i\|} \approx \frac{\|(\tilde{A} - E_i\,\tilde{B})\,\mathbf{c}\|}{\|\tilde{B}\,\mathbf{c}\|},$$

(20)

computed at the validation points. To determine the parameters, one can minimize different measures based on the residuals $r_i$. We found most useful to minimize

$$J = r_{max},$$

(21)

where $r_{max}$ is the maximum residual of all the desired $n_{\mathrm{lev}}$ energy levels.

To optimize the basis parameters, we first set $\beta_k$ equal to of $\beta_0$ and scan $V_{max}$, $V'_{max}$, and $\beta_0$, to determine optimal values. Second, $\alpha_k$ is optimized, where $\beta_k = \beta_0\ \alpha_k$. $\alpha_k$ is an anisotropic adjustment that was refined by coordinate descent. For a single coordinate, the optimization is done by comparing $J$ at the current value of $\alpha_k$ and at values that are 1% larger and smaller. The $M/N$ ratio is increased until $J$ and the levels no longer change appreciably. As reference energies are not used in this procedure, it is applicable when no reference spectrum is available.

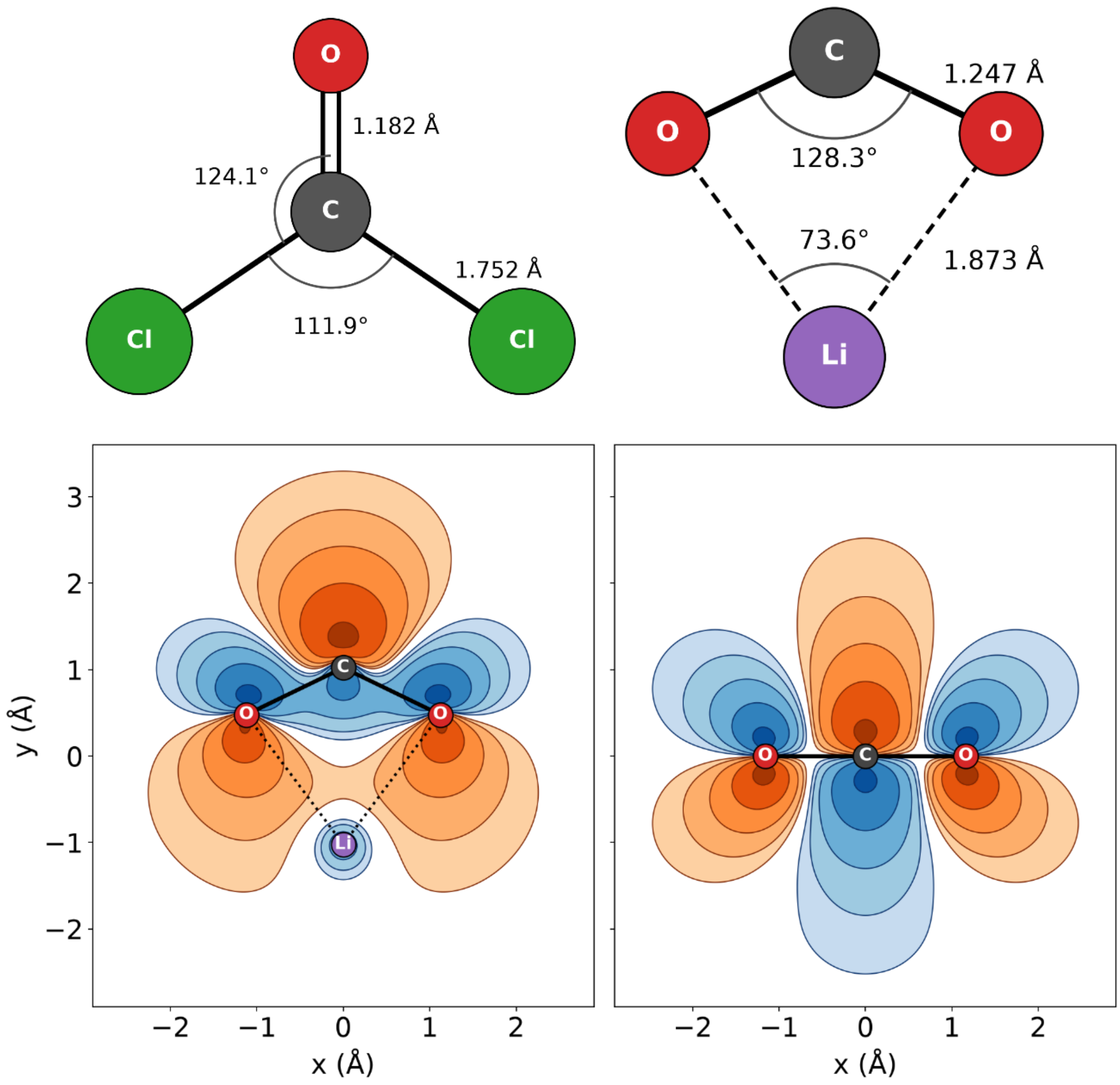

Figure 1. Molecules studied in this work: $Cl_2CO$ is planar in its equilibrium geometry on the PES of Ref. [34] (top left), and $LiCO_2$ is planar at its UB3LYP/aug-cc-pVTZ equilibrium geometry (top right). The bottom panels compare the SOMO of $LiCO_2$ (bottom left) and the LUMO of free $CO_2$ (bottom right).

# 3 Results

## *3.1 $Cl_2CO$*

### 3.1.1 Spectrum computed with a PES: benchmarking

We first compute the spectrum of phosgene on the PES of Ref. 34. The planar equilibrium geometry (Figure 1) on the PES is: r(C=O) = 1.18160 Å, r(C–Cl) = 1.75184 Å, ∠Cl-C-Cl = 111.851$^0$, ∠O-C-Cl = 124.075$^0$. The harmonic wavenumbers on the PES are: $\nu_1$ (C=O stretch) = 1851.72 cm$^{-1}$, $\nu_2$ (C–Cl symmetric stretch) = 572.99 cm$^{-1}$, $\nu_3$ (Cl–C–Cl bend) = 301.74 cm$^{-1}$, $\nu_4$ (out-of-plane) = 587.28 cm$^{-1}$, $\nu_5$ (C–Cl antisymmetric stretch) = 861.04 cm$^{-1}$, $\nu_6$ (in-plane rock) = 444.93 cm$^{-1}$. We use masses 12.000000 u for $^{12}$C, 15.994915 u for $^{16}$O, and 34.968853 u for $^{35}$Cl. Errors, relative to reference values from Ref. 34, are plotted in Figure 2 as a function of $N$. There are (blue) points for calculations done using only energies and (orange) points for calculations done with both energies and potential derivatives. Sufficiently large numbers of collocation points $M$ were used (as detailed in the figure caption). For the calculation with the derivatives, $p = 3$ was found to be best and was used for all $N$. The widths $\beta_k$ = {0.864, 0.971, 0.924, 0.960, 1.031, 0.936} (ordered by frequency: Cl-C-Cl bend, rocking, symmetric C–Cl stretch, the out-of-plane vibration, asymmetric C–Cl stretch, and 1851 C=O stretch) were optimized with $N$ = 5000 and $M$ = 15000 and using only energy rows (common width scanned from 0.80 to 1.00 in steps of 0.05, then coordinate descent with 5% and then 1% steps). They were used for all $N$ as their re-optimization was not found to be important. All optimizations are done for 100 energies.

The optimal $V_{max}$ was 5000 cm$^{-1}$ for $N$ = 2000 and 5000; and 5500 cm$^{-1}$ for $N$ = 10000. The optimal $V'_{max}$ was 1150 cm$^{-1}$ (≈ 0.5 × harmonic ZPE) in all cases. Note that these potential energy cutoff values are relatively low: reference values [34] for the ZPE and the 100$^{th}$ vibrational level are 2299.669 and 4358.031 cm$^{-1}$, respectively. As stressed before, accurate energies can be obtained using collocation even when it is not possible to compute integrals with a quadrature whose points are the collocation points.[49,53,65] This makes it possible to place points in a region of configuration space smaller than that required when using quadrature.[62,63]

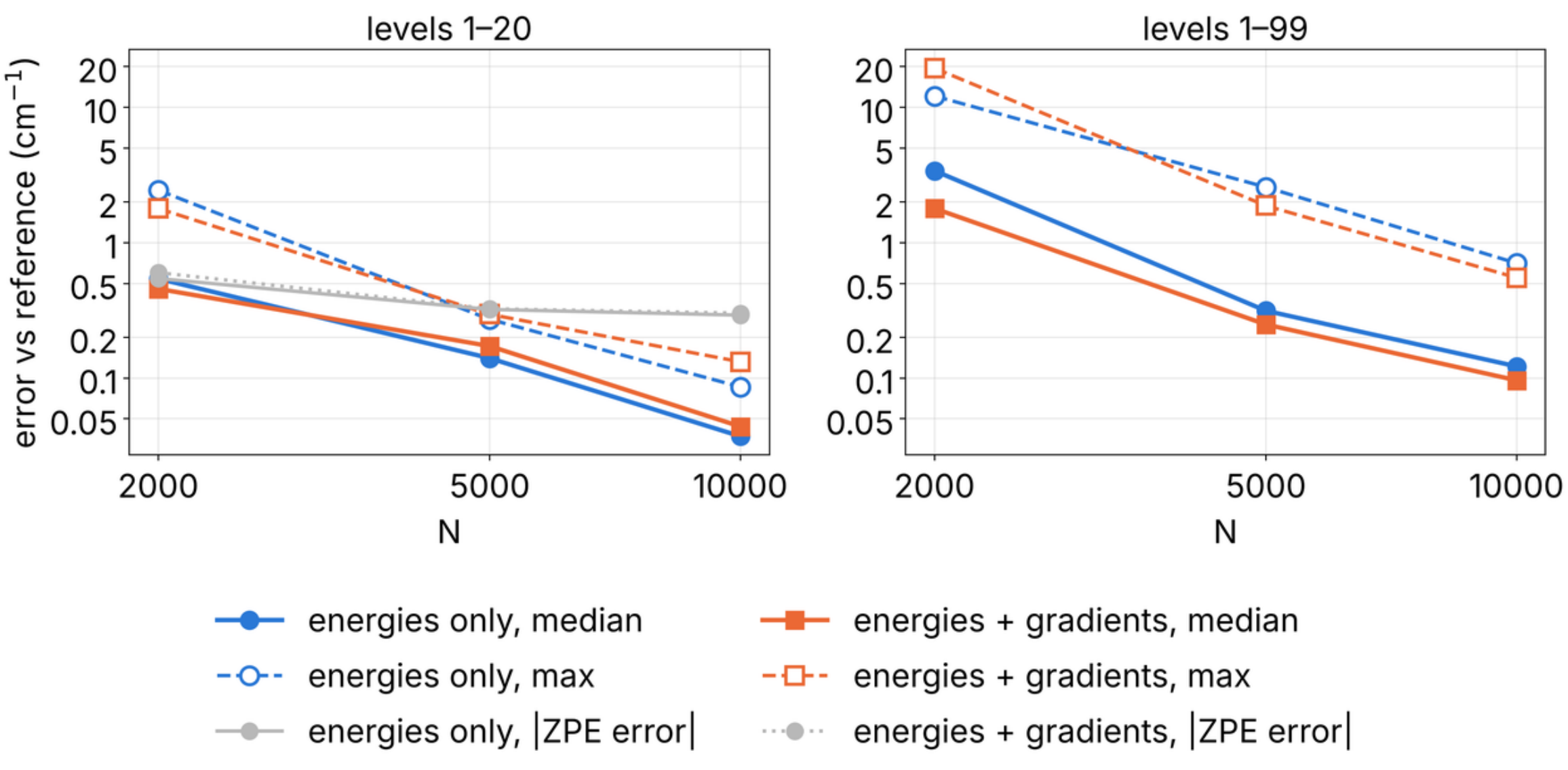

Figure 2. Errors with respect to the reference energies of Ref. [34], for the first 20 and first 99 vibrational levels, as well as ZPE, of $Cl_2CO$ as a function of the number $N$ of basis

functions. With energies only, $M/N = 6$ for $N = 2000$ and 5000, and $M/N = 4$ for $N = 10000$. When using derivatives, $M/N = 3$ for $N = 2000$ and 5000, and $M/N = 1$ for $N = 10000$.

One conclusion from these tests is that the achievable accuracy is limited by the basis size and cannot be improved by using potential derivatives. 5000 basis functions are sufficient to achieve sub-cm$^{-1}$ accuracy for most levels, and 10000 basis functions for all levels. This level of accuracy is sufficient for most materials science applications. It is less than the error due to the ab initio method. With 5000 or 10000 basis functions, the CPU and memory cost of solving the matrix eigenvalue problem is unproblematic on most personal computers. The cost of computing the matrices is also insignificant, when using a PES. However, when computing the matrices from ab initio energies, the cost of computing the matrices will scale with $M$. The value of $M$ necessary for obtaining sub-cm$^{-1}$ accuracy of the spectrum calculator without a PES is therefore a key factor for enabling calculations of anharmonic spectra directly from ab initio data.

Figure 3 shows error as a function of the $M/N$ ratio for different $N$. Overall, the use of derivatives makes it possible to use 4-6 times fewer PES (or ab initio) evaluations. The optimal Gaussian widths and $V'_{max}$ were not noticeably changed by adding gradient rows. At $N = 10000$, energies obtained with derivatives with 10000 points have smaller errors than those obtained without derivatives with 4 times more points: the calculation reaches a median error of 0.096 cm$^{-1}$ and a maximum error of 0.55 cm$^{-1}$ for levels 1–99, while the potential energy-only calculation with 40000 points gives 0.122 / 0.71 cm$^{-1}$ (median / max error). When using derivatives, it is possible to use fewer collocation points than basis functions. With 5000 or 10000 basis functions, the median errors of levels 1–19 with $M = 0.5N$ are 0.29 and 0.06 cm$^{-1}$ ($N = 5000$ and 10000), compared with 0.14 and 0.03 cm$^{-1}$ for the energy-only calculations with the largest $M$; with $M = N$, the median errors of levels 1–99 (0.32 and 0.10 cm$^{-1}$) are about the same as or smaller than those of the energy-only calculations with the largest $M$ (0.31 and 0.12 cm$^{-1}$). It is possible to compute all of the first 20 and most of the first 99 levels with an accuracy better than 1 cm$^{-1}$, using only 5000 basis functions and 2500-5000 collocation points (plus 2000 validation points used to monitor convergence). For comparison, in the Smolyak collocation calculation of Ref. [34], where 250 levels were computed, with smaller errors than those obtained here, about 2 million basis functions and collocation points were used.

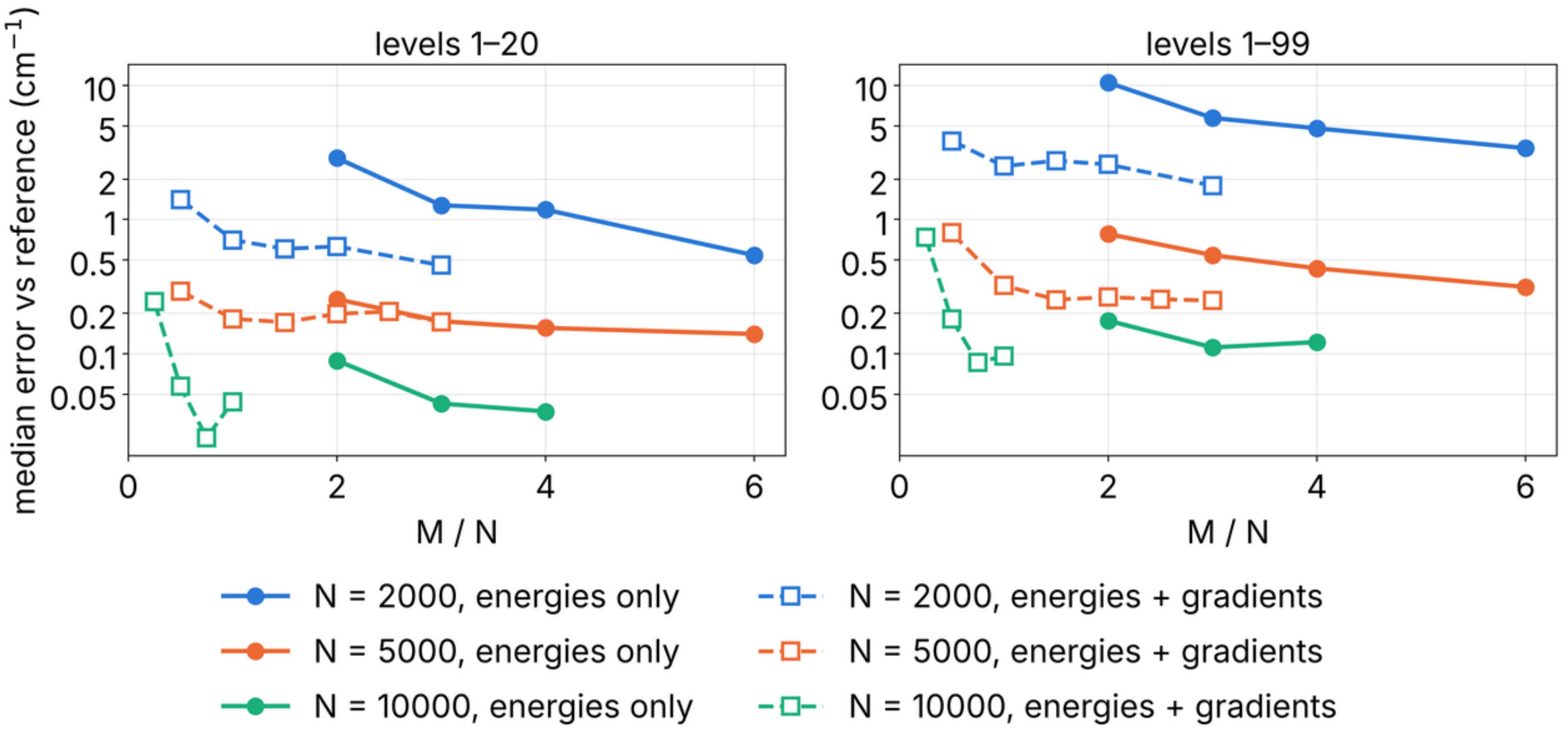

Figure 3. Errors with respect to reference energies for the first 20 and first 99 vibrational levels of $Cl_2CO$ as a function of *M/N*, the ratio of the number of collocation points to the number *N* of basis functions.

Figure 4 shows spectra (with linear and logarithmic scales) computed with the linear dipole approximation, from ab initio calculations at the equilibrium geometry. $N$ = 5000 and $M$ = 15000 were used. We use DFT/B3LYP with aug-cc-pVDZ and aug-cc-pVTZ basis sets and CCSD(T) with the frozen core approximation and the same basis sets to compute the dipole moment. In the CCSD(T) case, it is obtained with a field step size of ±0.001 a.u. In the DFT case, the dipole is output by PySCF. The dipole derivative is computed with a step of 0.02 a.u., along $\boldsymbol{q}$. We compare to $M$ = 5000 for B3LYP/aug-cc-pVDZ. The spectra agree for most practical purposes. One of the more recent spectra of phosgene is reported in Ref. 81, which we reproduce here and overlay on the top panel of Figure 4. Considering the vicissitudes of experimental measurements (such as the possibility of presence of other species) and the level of theory used for the dipole, the agreement between theory and experiment is good.

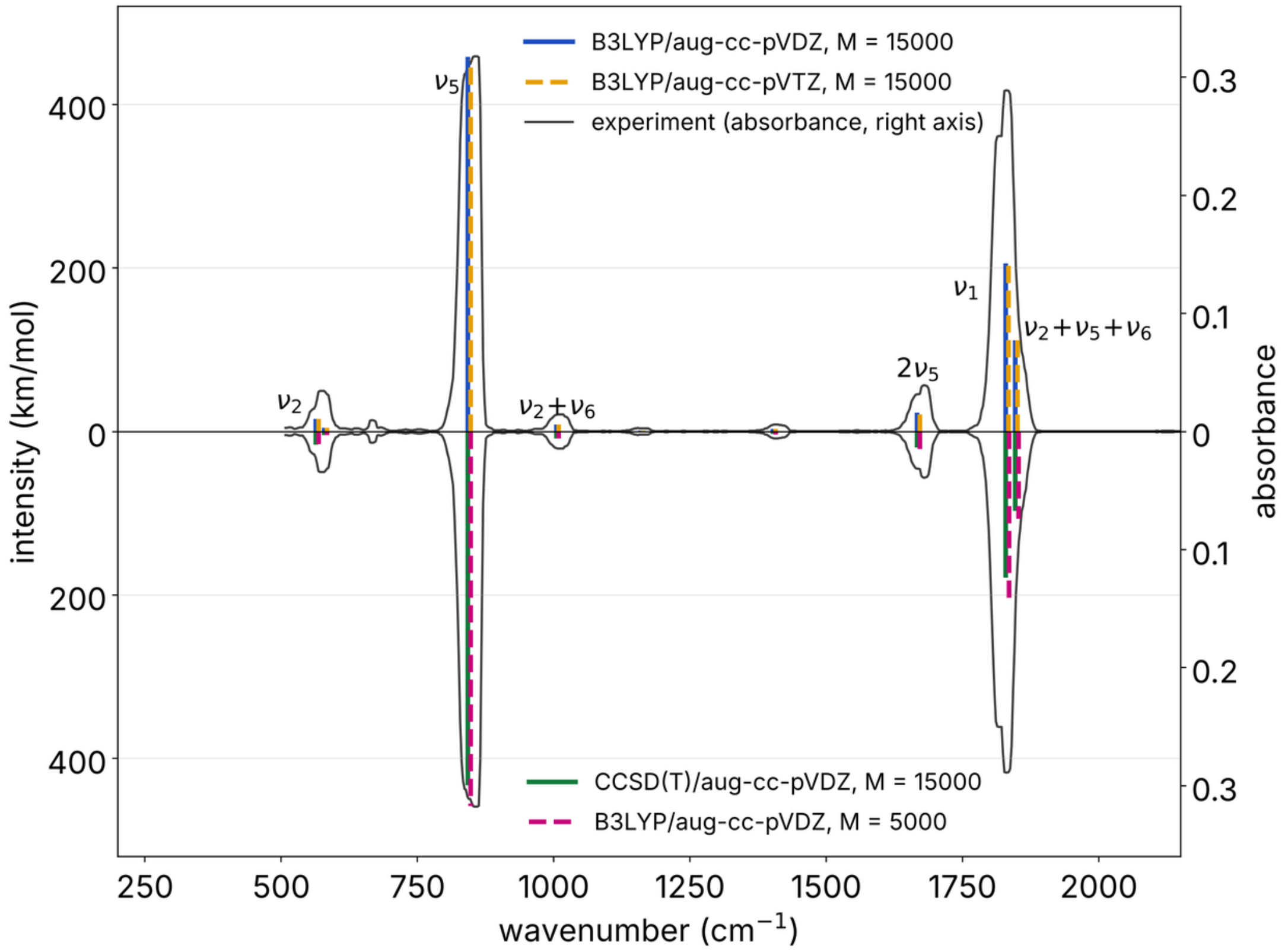
B3LYP/aug-cc-pVDZ, M = 15000
B3LYP/aug-cc-pVTZ, M = 15000
experiment (absorbance, right axis)
CCSD(T)/aug-cc-pVDZ, M = 15000
B3LYP/aug-cc-pVDZ, M = 5000
$\nu_5$
$\nu_2$
$\nu_2+\nu_6$
$2\nu_5$
$\nu_1$
$\nu_2+\nu_5+\nu_6$
intensity (km/mol)
absorbance
wavenumber (cm$^{-1}$)

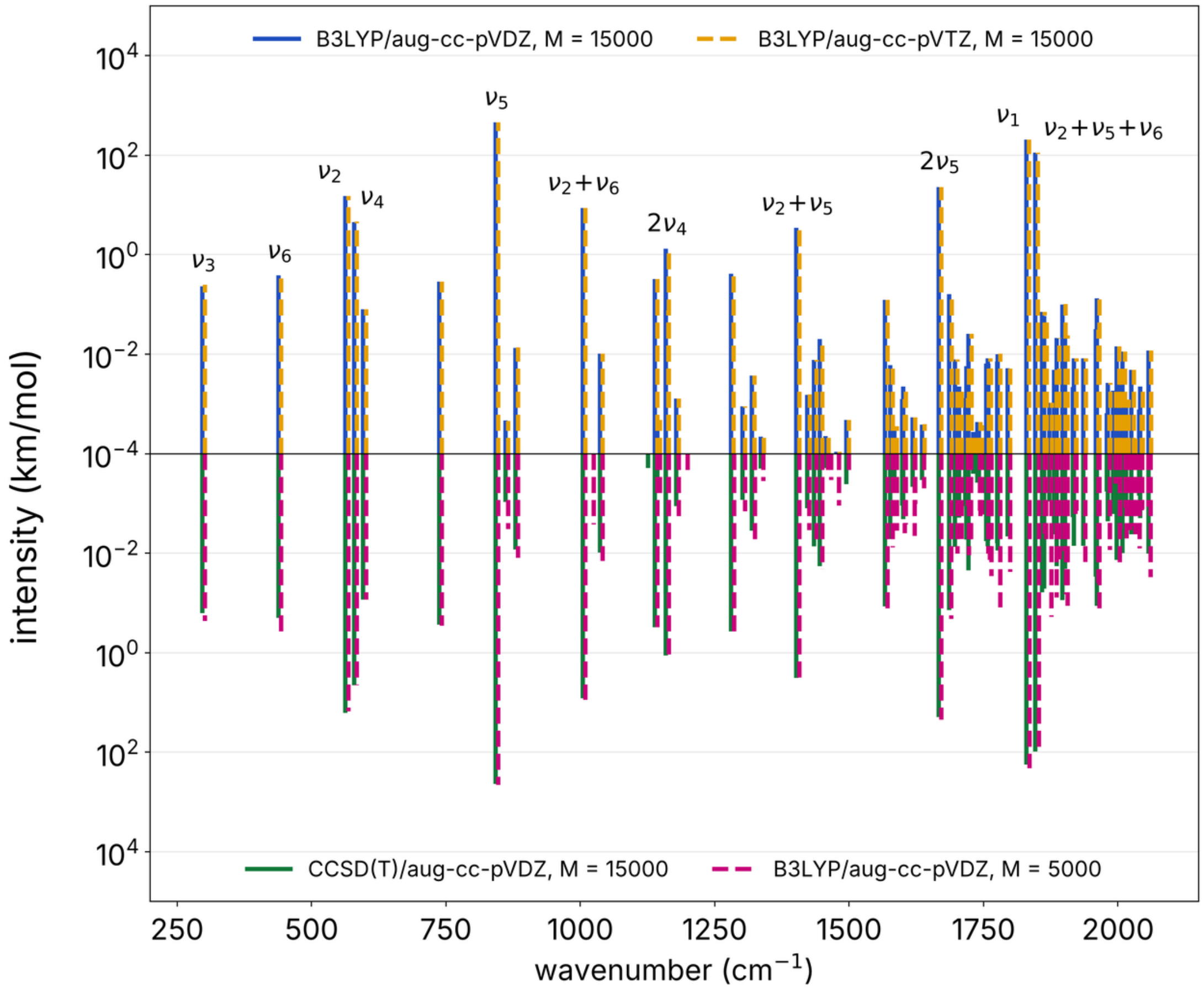
B3LYP/aug-cc-pVDZ, M = 15000
B3LYP/aug-cc-pVTZ, M = 15000
CCSD(T)/aug-cc-pVDZ, M = 15000
B3LYP/aug-cc-pVDZ, M = 5000
$\nu_3$
$\nu_6$
$\nu_2$
$\nu_4$
$\nu_5$
$\nu_2+\nu_6$
$2\nu_4$
$\nu_2+\nu_5$
$2\nu_5$
$\nu_1$
$\nu_2+\nu_5+\nu_6$
intensity (km/mol)
wavenumber (cm$^{-1}$)

Figure 4. Vibrational spectra (linear and logarithmic scales in the top and bottom panels, respectively) of $Cl_2CO$ computed on the PES and in the linear dipole approximation, with the dipole components computed at different levels of theory. Computed sticks are displaced horizontally for visibility. Above the zero line, B3LYP/aug-cc-pVDZ ($M$ = 15000) is shifted by −2.5 cm$^{-1}$ and B3LYP/aug-cc-pVTZ ($M$ = 15000) by +2.5 cm$^{-1}$. Below it, CCSD(T)/aug-cc-pVDZ ($M$ = 15000) is shifted by −2.5 cm$^{-1}$ and B3LYP/aug-cc-pVDZ ($M$ = 5000) by +2.5 cm$^{-1}$. The three $M$ = 15000 spectra have the same transition wavenumbers, and the $M$ = 5000 positions differ from them by less than 2 cm$^{-1}$. The experimental spectrum of Ref. 81 is included on the top panel, reproduced from Ref. 81 under CC BY license, © 2020 The Authors.

A summary of the accuracy obtained with different numbers of basis functions and collocation points, using potential energies only or energies and gradients, is given in Table 1. A full list of level energies is provided in Supporting Information.

Table 1. A summary of the accuracy of vibrational levels of $Cl_2CO$ computed on the PES, vs. reference calculations of Ref. 34, obtained with different numbers of basis functions $N$ and collocation points $M$, using potential energies only ($V$) or energies and gradients ($\nabla V$). The case "$p = 0$" corresponds to no weighing of gradient rows. The other results are for $p = 3$.

| **Use of $V, \nabla V$** | $N$ | $M$ | **Median / max error, levels 1–19** | **Median / max error, levels 1–99** |
|---|---|---|---|---|
| $V$ | 5000 | 30000 | 0.14 / 0.27 | 0.31 / 2.57 |
| $V$ | 10000 | 30000 | 0.04 / 0.13 | 0.11 / 0.73 |
| $V$ | 10000 | 40000 | 0.03 / 0.09 | 0.12 / 0.71 |
| $V$ and $\nabla V$ | 5000 | 2500 | 0.29 / 0.78 | 0.79 / 7.75 |
| $V$ and $\nabla V$ | 5000 | 5000 | 0.19 / 0.35 | 0.32 / 2.91 |
| $V$ and $\nabla V$ | 5000 | 15000 | 0.18 / 0.30 | 0.25 / 1.88 |
| $V$ and $\nabla V$ | 10000 | 5000 | 0.06 / 0.30 | 0.18 / 1.04 |
| $V$ and $\nabla V$ | 10000 | 7500 | 0.02 / 0.15 | 0.09 / 0.78 |
| $V$ and $\nabla V$ | 10000 | 10000 | 0.04 / 0.13 | 0.10 / 0.55 |
| $V$ and $\nabla V$, $p = 0$ | 10000 | 10000 | 0.02 / 0.15 | 0.19 / 1.74 |

### 3.1.2 Spectrum computed without a PES

The ability to compute a spectrum from a relatively small number of potential points makes the method attractive for cases where there is no PES. This includes many materials science applications, in which the feasible ab initio method is typically DFT. The collocation spectra (potential gradients were used) for $Cl_2CO$, computed with DFT, B3LYP/aug-cc-pVDZ and B3LYP/aug-cc-pVTZ are shown in Figure 5. The widths $\beta_k$ = {0.831, 0.929, 0.978, 0.988, 1.068, 0.950} (same order of modes as on the PES) were optimized with the procedure of Section 2.3 ($J = r_{max}$) on the B3LYP/aug-cc-pVDZ data, with $N = M = 5000$ with gradient rows ($p = 3$, $V_{max} = 5000$ cm$^{-1}$, $V_{max}' = 1150$ cm$^{-1}$), with 1000 validation points (common width scanned from 0.80 to 1.00 in steps of 0.05, then coordinate descent with 5% and then 1% steps), and the same widths were used for B3LYP/aug-cc-pVTZ. The dipole moment surface used for the intensities was a polynomial, of total degree 4, in the six normal coordinates, fitted to DFT dipole moment values computed at the collocation points. The error of the dipole moment was less than $1\times10^{-4}$ au. at the validation points, for every component. Equilibrium geometries at these levels of theory are: r(C=O) = 1.1801 Å, r(C–Cl) = 1.7668 Å, ∠(Cl–C–Cl) = 111.86° with aug-cc-pVDZ, and r(C=O) = 1.1734 Å, r(C–Cl) = 1.7602 Å, ∠(Cl–C–Cl) = 111.82° with aug-cc-pVTZ. The harmonic wavenumbers are: $\nu_1$ (C=O stretch) = 1862.36 cm$^{-1}$, $\nu_2$ (C–Cl symmetric stretch) = 556.67 cm$^{-1}$, $\nu_3$ (Cl–C–Cl bend) = 297.10 cm$^{-1}$, $\nu_4$ (out-of-plane) = 579.23 cm$^{-1}$, $\nu_5$ (C–Cl antisymmetric stretch) = 818.73 cm$^{-1}$, $\nu_6$ (in-plane rock) = 432.86 cm$^{-1}$ with aug-cc-pVDZ, and $\nu_1$ (C=O stretch) = 1866.68 cm$^{-1}$, $\nu_2$ (C–Cl symmetric stretch) = 554.29 cm$^{-1}$, $\nu_3$ (Cl–C–Cl bend) = 296.09 cm$^{-1}$, ν4 (out-of-plane) = 585.85 cm$^{-1}$, $\nu_5$ (C–Cl antisymmetric stretch) = 809.76 cm$^{-1}$, $\nu_6$ (in-plane rock) = 434.94 cm$^{-1}$ with aug-cc-pVTZ. There are no significant differences between aug-cc-pVDZ and aug-cc-pVTZ spectra, but both noticeably deviate from the spectrum computed on the spectroscopic-quality PES, due to a lower level of theory (the PES was fitted to CCSD(T)/cc-pVTZ data[34]).

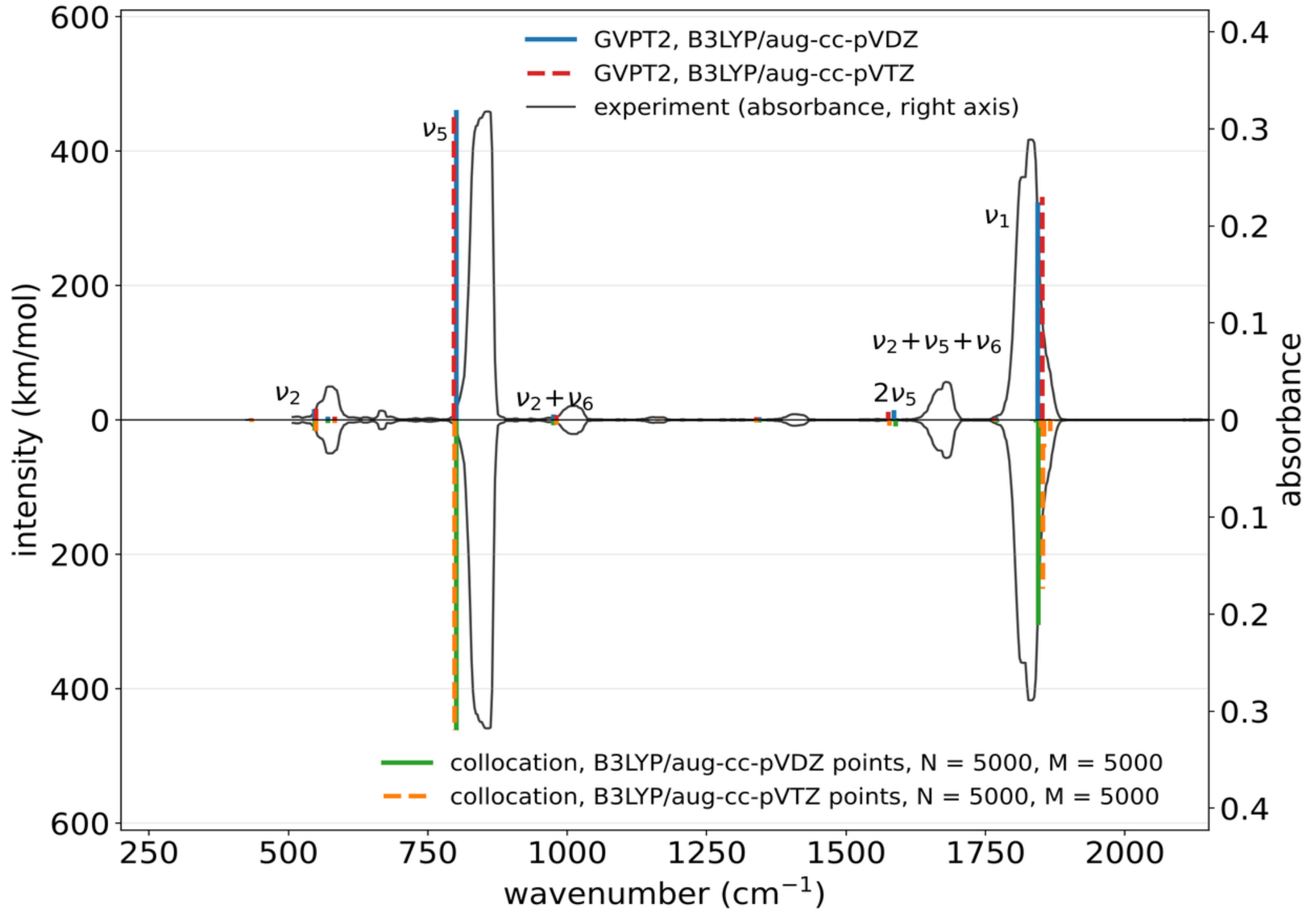

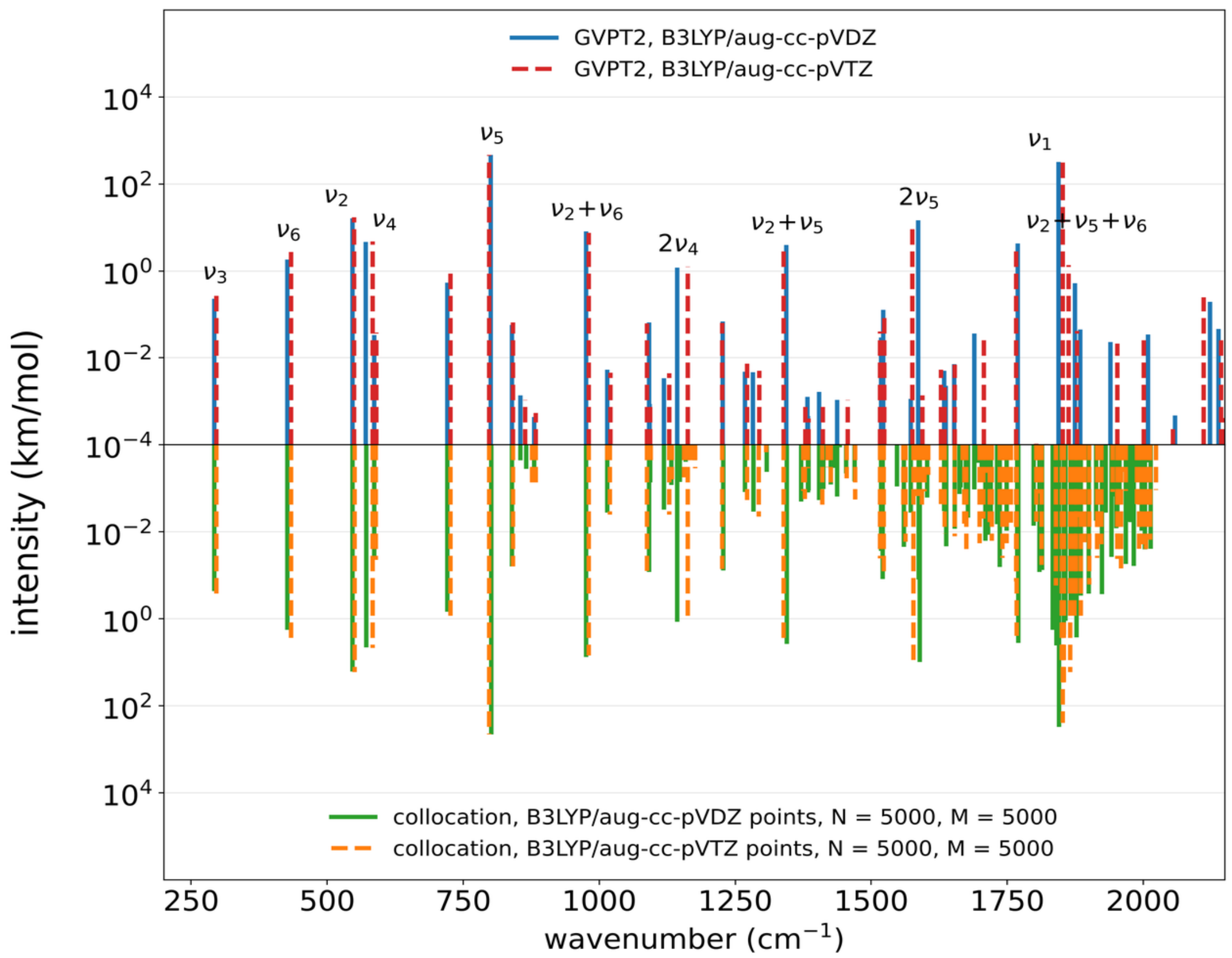

Figure 5. Vibrational spectra (linear and logarithmic scales in the top and bottom panels, respectively) of $Cl_2CO$ computed with energies using DFT/B3LYP with aug-cc-pVDZ and -pVTZ bases (*N*, *M* = 5000), as well as spectra computed with GVPT2 at the same

levels of ab initio theory. For better readability, the lines with aug-cc-pVDZ are shifted by −2.5 cm$^{-1}$ and those with aug-cc-pVTZ by +2.5 cm$^{-1}$. The experimental spectrum of Ref. 81 is included on the top panel, reproduced from Ref. 81 under CC BY license, © 2020 The Authors.

Figure 5 also compares with the GVPT2[15,18,19] spectrum computed from force constants calculated at the same level of ab initio theory. The GVPT2 calculation takes account of the resonance due to the proximity of $\nu_1$ and $\nu_2$+$\nu_5$+$\nu_6$ levels. It uses more quartic constants than the standard semi-diagonal set; the force field has 207 terms, including 6 diagonal $f_{iiii}$, 45 two-mode $f_{iijj}$ and $f_{iiij}$, 60 three mode $f_{iijk}$, and four-mode $f_{1256}$. The π- π term in the Watson KEO was included and the effective moment of inertia tensor is evaluated at equilibrium. The collocation and GVPT2 spectra agree well in this case, as expected, as it is relatively harmonic. Representative frequences and intensities, those for transitions with significant intensities, with both collocation and GVPT2 are listed in Table 2, where they are also compared to the calculations on the PES. The collocation spectrum is denser than the VPT2 spectrum because it is built from the 100 lowest levels, regardless of their character (highest excitation included has 7 quanta), while the GVPT2 spectrum is limited to 4 quanta. Another reason is intensity borrowing. All the collocation states mix with $\nu_1$, $\nu_5$ and the other bright states and therefore acquire intensity. On the other hand, because GVPT2 uses a quartic force field, there is less mixing. The full list of level energies is provided in Supporting Information.

Table 2. Comparison of representative transition wavenumbers (cm$^{-1}$) / intensities (km mol$^{-1}$) of $Cl_2CO$ computed with different methods. Collocation calculation results are given for *N, M* = 5000 (using potential derivatives). "aDZ" and "aTZ" signify B3LYP/aug-cc-pVDZ and B3LYP/aug-cc-pVDZ, respectively.

| State | Collocation | Collocation, DFT | | GVPT2 | |
|---|---|---|---|---|---|
| | PES | aDZ | aTZ | aDZ | aTZ |
| $\nu_3$ | 299.43 / 0.25 | 294.55 / 0.23 | 293.72 / 0.29 | 294.44 / 0.23 | 293.68 / 0.29 |
| $\nu_6$ | 441.14 / 0.33 | 429.25 / 1.78 | 431.29 / 2.73 | 429.17 / 1.80 | 431.23 / 2.73 |
| $\nu_2$ | 565.94 / 15.2 | 549.43 / 16.2 | 547.40 / 16.9 | 549.31 / 16.3 | 547.24 / 16.9 |
| $\nu_4$ | 581.51 / 4.63 | 573.82 / 4.57 | 580.91 / 4.72 | 573.73 / 4.58 | 580.92 / 4.73 |

| State | Collocation | Collocation, DFT | | GVPT2 | |
|---|---|---|---|---|---|
| | PES | aDZ | aTZ | aDZ | aTZ |
| $\nu_3+\nu_6$ | 740.33 / 0.28 | 723.17 / 0.68 | 724.29 / 1.07 | 723.26 / 0.54 | 724.49 / 0.86 |
| $\nu_5$ | 844.86 / 455.0 | 803.68 / 462.1 | 795.18 / 462.1 | 803.15 / 461.5 | 794.86 / 461.6 |
| $\nu_6+\nu_2$ | 1006.95 / 8.77 | 978.24 / 7.59 | 978.15 / 7.01 | 978.19 / 8.13 | 978.03 / 7.48 |
| $2\nu_4$ | 1162.02 / 1.35 | 1146.27 / 1.16 | 1160.67 / 1.20 | 1146.08 / 1.19 | 1160.76 / 1.25 |
| $\nu_2+\nu_5$ | 1404.80 / 3.43 | 1346.99 / 3.84 | 1336.89 / 3.54 | 1346.43 / 3.99 | 1336.60 / 3.76 |
| $2\nu_5$ | 1669.37 / 22.6 | 1591.27 / 9.83 | 1574.81 / 8.78 | 1588.35 / 14.6 | 1572.70 / 12.4 |
| $\nu_2+\nu_5+\nu_6$ | 1850.45 / 107.4 | 1772.35 / 3.59 | 1764.89 / 2.48 | 1771.39 / 4.31 | 1763.83 / 3.12 |
| $\nu_1$ | 1832.90 / 210.6 | 1847.71 / 305.2 | 1849.91 / 251.4 | 1846.41 / 323.8 | 1849.76 / 332.1 |
| $3\nu_3+\nu_6+\nu_2$ | 1898.19 / 0.01 | 1856.25 / 2.3 | 1851.99 / 40.9 | not matched | not matched |
| $\nu_3+2\nu_5$ | 1963.11 / 0.16 | 1880.00 / 2.7 | 1863.65 / 16.8 | 1876.63 / 0.5 | 1860.42 / 1.3 |

### 3.2 *$LiCO_2$*

We now consider a much more anharmonic case: $LiCO_2$. The equilibrium geometry is shown in Figure 1. We choose this molecule because in it the donation of charge from Li to the $CO_2$ LUMO-derived state (as can also be appreciated from Figure 1) is similar to that of a catalytic surface in many molecule-surface systems, for the majority of which there is no PES function. Catalytic bond weakening makes the inclusion of anharmonic effects more important than in a free molecule, especially when vibrational excitations involve several quanta. For many catalytic systems, DFT is often the only feasible approach. The weakly bound, charge-transfer nature of $LiCO_2$ poses challenges to VPT and other methods that rely on a polynomial expansion of the PES near equilibrium (such as VCI[25,82,83] that is often used with such a PES). For example, anharmonic force constants obtained by finite differences of analytic DFT Hessians proved highly sensitive to integration-grid errors. Because the analytic Hessian computed by PySCF does not use grid-weight derivatives (to account for the effect of grid changes when atoms are displaced), one semi-diagonal quartic constant of $CO_2$ varied by more than a factor of two (−111 vs −236 cm$^{-1}$) depending on the which derivatives are those in the Hessian and which are done by finite difference, shifting VPT2 fundamentals by up to ~20 cm$^{-1}$ at the default grid, even though the harmonic frequencies were essentially unaffected. Consistent results required a much denser grid (PySCF level 8). Also, we found that the quartic force field of this molecule had a hole when the ~300 cm$^{-1}$ mode corresponding to Li displacement was increased and the other coordinates were relaxed. Along this cut the potential has a barrier of only about 1316 cm$^{-1}$ at $q \approx 3$ and then decreases to negative values. States with three or more quanta in that mode (about 910 cm$^{-1}$) are near or above that barrier and are not true bound states. These two problems demonstrate that it is desirable to go beyond potentials computed from equilibrium geometry derivatives.

Collocation calculations of the lowest 100 vibrational levels were done with potential derivatives, and $p = 0$ was found to be best for this floppier molecule. The Gaussian center cutoff $V_{\max}'$, $V_{\max}$, $p$ and the Gaussian widths were tuned with the UB3LYP-D3(BJ)/aug-cc-pVDZ collocation points using the procedure of Section 2.3 ($J = r_{\max}$) and $N = 10000$ and $M = 12000$, with 1275 validation points: $V_{\max}'$ was scanned from 700 to 1500 cm$^{-1}$, then $V_{\max} = 5000$ and 5500 cm$^{-1}$ and $p = 0$, 1 and 3 were compared, the common width was scanned in steps of 0.05, and the widths were then optimized by coordinate descent with 5% and then 1% steps. The same parameters were used for aug-cc-pVTZ. $V_{max}$ and $V_{max}'$ were 5000 and 1200 cm$^{-1}$, respectively. Gaussian basis function widths in dimensionless normal coordinates, ν1 - ν6 were 0.9321, 0.99, 1.00, 1.0188, 1.00, and 1.00, respectively. The spectra computed with collocation on the UB3LYP(BJ)/aug-cc-pVDZ, -pVTZ potentials are shown in Figure 6. They are also compared to GVPT2. $N = 14000$ and $M$ of about 12000 were sufficient to converge the first 19 levels to better than 1 cm$^{-1}$. Of all levels, 57 (aDZ) and 54 (aTZ) levels were converged within 1 cm$^{-1}$, and 70 within 2 cm$^{-1}$. The 11 levels below 1000 cm$^{-1}$, which include ν1–ν4 and 2ν2, are converged to 0.3 cm$^{-1}$ or better. The convergence error was estimated as the range of values for each level over four calculations: $N = 10000$ and 14000, each with two independent sets of DFT points. All levels are given in Supporting Information.

Rectangular collocation and GVPT2 agree well for the fundamentals: they differ by at most 2 cm$^{-1}$ for aug-cc-pVTZ and 5 cm$^{-1}$ at aug-cc-pVDZ. Their IR intensities agree to within a few percent for the strong bands ν1, ν2, ν3 and ν6. As expected, differences become more substantial for overtones and combination bands. For the clearly assigned states (states that can be identified with a specific harmonic-oscillator state) up to 2200 cm$^{-1}$ (about 60), the mean absolute deviation is 5–8 cm$^{-1}$ and the RMS deviation 7–11 cm$^{-1}$. There are levels that differ by up to 20–30 $cm^{-1}$, mainly overtones and combinations of the low-frequency modes involving Li, for example 2ν3 (for which the frequency with collocation is higher by 18 $cm^{-1}$ at aDZ and by 10 $cm^{-1}$ at aTZ) and 2ν1 (lower by 2–6 $cm^{-1}$ with collocation). These deviations are an order of magnitude larger than the collocation convergence error (0.3 $cm^{-1}$ below 1000 $cm^{-1}$, 2.4–2.9 $cm^{-1}$ RMS over all 99 levels), so they likely reflect the limits of the quartic force field and second-order perturbation treatment. The two methods also result in different intensity sharing among near-degenerate states, for example, the intensity ratio within the ν2 + ν5 pair near 1650–1680 $cm^{-1}$.

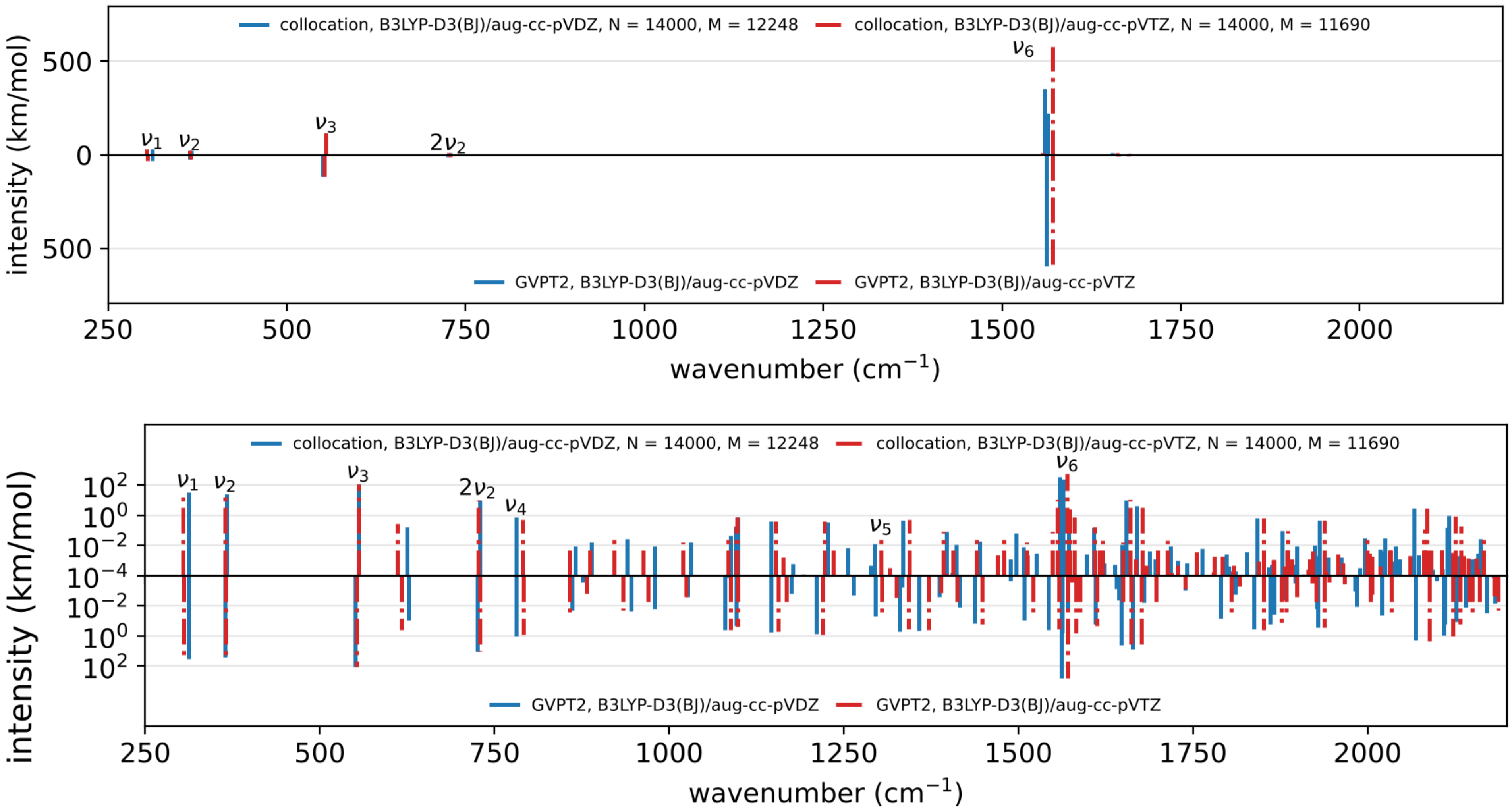

Figure 6. The spectra of $LiCO_2$ computed with collocation and GVPT2 on B3LYP(BJ)/aug-cc-pVDZ, -pVTZ potentials, linear (top) and log (bottom) scales.

Increasing the DFT basis size from aug-cc-pVDZ to aug-cc-pVTZ shifts the collocation levels by about 10 $cm^{-1}$ on average (MAE 10 $cm^{-1}$, RMS 12 $cm^{-1}$ for the clearly assigned states up to 2200 $cm^{-1}$, maximum about 27 $cm^{-1}$). The shift is systematic and follows the harmonic frequencies. The $CO_2$-moiety fundamentals ν4, ν5 and ν6 increase by 9–11 $cm^{-1}$, the Li out-of-plane mode ν1 drops by 8 $cm^{-1}$, and ν2 and ν3 barely move (−1.7 and 0.0 $cm^{-1}$). The harmonic wavenumbers change by +10.2, +10.1, +8.8, −10.5, −1.0 and +0.3 $cm^{-1}$ respectively, so the anharmonic contributions are nearly basis-independent. Accordingly, overtones and combinations with several ν1 quanta move down and those containing ν4–ν6 move up. Both basis sets calculations are converged to the same degree: the first 19 levels to 1 $cm^{-1}$ or better, and all 99 levels to an RMS of 2.9 $cm^{-1}$ at -pVDZ and 2.4 $cm^{-1}$ at -pVTZ (maximum 12 $cm^{-1}$ at -pVDZ, 8 $cm^{-1}$ at -pVTZ). The basis-set effect is therefore larger than both the convergence error and the GVPT2–collocation difference. The intensity patterns are similar, with the exception that ν6 mixes with a nearby B2 state (same symmetry as ν6) and its intensity splits 354:224 km/mol between levels about 4 $cm^{-1}$ apart. With the pVTZ basis, the partner keeps only 11 km/mol, giving a single ν6 band at 1570.1 $cm^{-1}$ that accidentally matches the experimental Ar-matrix value of 1568.6–1569.9 $cm^{-1}$.[84] The -pVTZ basis results are also closer to the experiment for ν4 (790 vs 799 $cm^{-1}$) and ν5 (1304 vs 1325–1330 $cm^{-1}$). This highlights the importance of a well-converged ab initio setup. Nevertheless, this is not critical for the purpose of demonstrating the performance of rectangular collocation method of this paper, which works with potential data computed at any level of theory, with or without derivatives. Some representative bands are summarized in Table 3.

Table 3. Vibrational band origins (cm$^{-1}$) and vibrational band intensities (km/mol, in parentheses) of $LiCO_2$ at the UB3LYP-D3(BJ) level with the aug-cc-pVDZ ("DZ") and aug-cc-pVTZ ("TZ") basis sets, computed with rectangular collocation and GVPT2, with some frequencies compared with experimental Ar-matrix values.[84]

| Band | Harmonic | | Collocation | | GVPT2 | | Ar matrix[c] |
|---|---|---|---|---|---|---|---|
| | DZ | TZ | DZ | TZ | DZ | TZ | |
| ν1 Li out-of-plane | 311.1 | 300.5 | 311.6 (31.3) | 303.7 (30.5) | 312.2 (31.4) | 305.2 (30.6) | — |
| ν2 Li in-plane | 388.2 | 387.2 | 366.2 (24.8) | 364.4 (24.9) | 364.0 (24.6) | 364.8 (24.6) | — |
| ν3 Li–$O_2$ stretch | 581.4 | 581.7 | 555.0 (118.5) | 555.0 (115.5) | 550.1 (117.3) | 553.0 (115.1) | 533.4, 526.0, 520.0 |
| ν4 OCO bend | 786.6 | 796.8 | 780.9 (0.74) | 790.1 (0.52) | 781.5 (1.05) | 791.2 (0.87) | 798.7 |
| ν5 sym. CO stretch | 1322.3 | 1332.3 | 1294.2 (0.014) | 1303.5 (0.035) | 1295.4 (0.048) | 1304.8 (0.070) | 1329.9, 1325.0 |
| ν6 asym. CO stretch | 1600.9 | 1609.7 | 1559.4 (354), 1563.5 (224) | 1570.1 (580) | 1562.0 (596) | 1571.0 (611) | 1568.6, 1569.9 |
| 2ν1 | 622.2 | 601.1 | 625.0 (0.18) | 611.1 (0.26) | 627.4 (0.09) | 616.7 (0.39) | — |
| 2ν2 | 776.4 | 774.4 | 728.9 (9.3) | 726.2 (9.9) | 724.9 (10.5) | 728.5 (10.4) | — |
| 2ν3 | 1162.8 | 1163.4 | 1097.9 (0.78) | 1097.4 (0.76) | 1079.8 (0.42) | 1087.8 (0.38) | — |
| 2ν4 | 1573.3 | 1593.6 | 1561.4 (0.64) | 1580.2 (0.74) | 1563.1 (0.61) | 1582.4 (0.63) | — |
| ν4 + ν5 | 2108.9 | 2129.1 | 2066.5 (2.7) | 2084.7 (2.8) | 2068.6 (2.0) | 2087.9 (2.3) | — |

## 4 Conclusion

In this paper we modify and demonstrate the utility of a rectangular collocation method that uses the exact KEO in space-fixed Cartesian coordinate. Although it is simple, the KEO is exact, no coupling (e.g. the π- π term in the Watson KEO) is neglected. Variational calculations are always hampered by the need to evaluate the PES at a huge number of (quadrature) points. We show that by using Claude to optimize the parameters of a Gaussian basis, it is possible to obtain a basis good enough that, despite the large condition number of $\boldsymbol{B}$ (and the associated large Lebesgue constant[71]), one can compute an accurate spectrum with a small basis and a small point set.

The approach of Ref. 51 is modified in two ways. (1) The calculation of vibrational intensities was implemented. This clearly demonstrates that not only energies computed by collocation but also the associated wavefunctions are accurate. (2) The new method uses potential derivatives. In a standard collocation calculation, one requires that wavefunctions (in a finite basis) exactly solve the Schrödinger equation at a set of points. Using derivatives make it possible to apply additional constraints and therefore to reduce the required number of points. From $M$ points one obtains $M(D+1)$ equations. Derivatives reduce the number of points needed by about a factor of 3–6 for the four-atom molecules considered here. This is most important if one wishes to compute a spectrum directly from ab initio data. It is popular to use various flavors of "direct dynamics" when using classical mechanics. When using quantum mechanics, the number of required quadrature points is usually so large that direct dynamics is too costly. For this reason, one invariably begins by fitting a PES. We have shown that a Gaussian basis can be simultaneously small enough that direct dynamics is possible and good enough to get accurate spectra. The size of the basis cannot be reduced by using derivatives. The basis must still be chosen so that desired wavefunctions are linear combinations of the basis functions.

The method has many advantages over traditional variational methods that require a molecule-fixed KEO and quadrature. Most importantly, because the basis size is small, the size of the required point set and the size of the final eigenvalue problem are small. Dozens of low-lying energy levels with errors on the order of 1 $cm^{-1}$ can be computed with only a few thousand PES points and basis function. This makes the method attractive for calculations of spectra directly from ab initio data, e.g. in materials science applications where a PES function is not available. The accuracy is limited by the accuracy of the underlying ab initio method rather than the accuracy of the solutions of the vibrational Schrödinger equation. Critical to the success of the method is the ability to assess the accuracy of energy levels by computing residuals for the rectangular eigenvalue problem from which the final (square) eigenvalue problem is obtained.

The entire procedure is taught to Claude and implemented as a Claude skill. In our experience, this is immensely helpful. Claude executes routine tasks that would otherwise be extremely time consuming for a human being. Examples include coordinate transformations, optimization of various parameters (such as basis parameters, energy cutoffs, numbers of basis functions and collocation points), and interfacing with ab initio codes. In the paper, we have presented ideas for using the space-fixed KEO with Gaussian basis functions that are functions of normal coordinates. However, calculations with other basis functions and other internal coordinates could be implemented in the same way by using the Claude Skill. It would also be possible to do calculations with a subset of the coordinates, as was demonstrated for acetic acid on titania in Ref. [41,60]. Overall, we believe the method is a general and versatile tool for numerically exact anharmonic vibrational spectroscopy. Especially nice is the fact that it can be used without a potential energy surface.

## 5 Supplementary Material

The supplementary material includes spectra computed with different methods.

## 6 Acknowledgements

Tucker Carrington thanks NSERC for financial support. Moksh Bhateja was supported by MITACS Globalink program.

## 7 Data availability statement

Energy levels computed in this work are provided in Supporting Information. The codes, data, and example of usage are also available as Claude Skill on https://github.com/sergeimanzhos/rectcolloc-vib .

## 8 References

[1] V. Barone, S. Alessandrini, M. Biczysko, J.R. Cheeseman, D.C. Clary, A.B. McCoy, R.J. DiRisio, F. Neese, M. Melosso, and C. Puzzarini, "Computational molecular spectroscopy," Nat. Rev. Methods Primers **1**(1), 38 (2021).

[2] M.J. Baker, J. Trevisan, P. Bassan, R. Bhargava, H.J. Butler, K.M. Dorling, P.R. Fielden, S.W. Fogarty, N.J. Fullwood, K.A. Heys, C. Hughes, P. Lasch, P.L. Martin-Hirsch, B. Obinaju, G.D. Sockalingum, J. Sulé-Suso, R.J. Strong, M.J. Walsh, B.R. Wood, P. Gardner, and F.L. Martin, "Using Fourier transform IR spectroscopy to analyze biological materials," Nat. Protoc. **9**(8), 1771–1791 (2014).

[3] F. Zaera, "New advances in the use of infrared absorption spectroscopy for the characterization of heterogeneous catalytic reactions," Chem. Soc. Rev. **43**(22), 7624–7663 (2014).

[4] C. Puzzarini, J. Bloino, N. Tasinato, and V. Barone, "Accuracy and Interpretability: The Devil and the Holy Grail. New Routes across Old Boundaries in Computational Spectroscopy," Chem. Rev. **119**(13), 8131–8191 (2019).

[5] V. Barone, M. Biczysko, and J. Bloino, "Fully anharmonic IR and Raman spectra of medium-size molecular systems: accuracy and interpretation," Phys. Chem. Chem. Phys. **16**(5), 1759–1787 (2014).

[6] J. Bloino, A. Baiardi, and M. Biczysko, "Aiming at an accurate prediction of vibrational and electronic spectra for medium-to-large molecules: An overview," International Journal of Quantum Chemistry **116**(21), 1543–1574 (2016).

[7] J.M. Bowman, T. Carrington, and H.-D. Meyer, "Variational quantum approaches for computing vibrational energies of polyatomic molecules," Mol. Phys. **106**(16–18), 2145–2182 (2008).

[8] P.R. Franke, J.F. Stanton, and G.E. Douberly, "How to VPT2: Accurate and Intuitive Simulations of CH Stretching Infrared Spectra Using VPT2+K with Large Effective Hamiltonian Resonance Treatments," J. Phys. Chem. A **125**(6), 1301–1324 (2021).

[9] M. Chan, K. Yamashita, T. Carrington, and S. Manzhos, "Towards Accurate Spectroscopic Identification of Species at Catalytic Surfaces: Anharmonic Vibrations of Formate on AuPt," MRS Proc. **1484**, (2012).

[10] S. Manzhos, T. Carrington, and K. Yamashita, "Calculating anharmonic vibrational frequencies of molecules adsorbed on surfaces directly from ab initio energies with a molecule-independent method: H2O on Pt(111)," Surf. Sci. **605**(5), 616–622 (2011).

[11] X. Huang, and T.J. Lee, "A procedure for computing accurate ab initio quartic force fields: Application to HO2+ and H2O," J. Chem. Phys. **129**(4), 044312 (2008).

[12] H. Rabitz, and Ö.F. Aliş, "General foundations of high-dimensional model representations," J. Math. Chem. **25**(2), 197–233 (1999).

[13] S. Manzhos, and M. Ihara, "Orders-of-coupling representation achieved with a single neural network with optimal neuron activation functions and without nonlinear parameter optimization," Artif. Intell. Chem. **1**(2), 100013 (2023).
[14] S. Carter, S.J. Culik, and J.M. Bowman, "Vibrational self-consistent field method for many-mode systems: A new approach and application to the vibrations of CO adsorbed on Cu(100)," J. Chem. Phys. **107**(24), 10458–10469 (1997).
[15] H.H. Nielsen, "The Vibration-Rotation Energies of Molecules," Rev. Mod. Phys. **23**(2), 90–136 (1951).
[16] I.M. Mills, "3.2 - Vibration–Rotation Structure in Asymmetric- and Symmetric-Top Molecules," in *Molecular Spectroscopy*, edited by K.N. Rao and C.W. Mathews, (Academic Press, 1972), pp. 115–140.
[17] E.L. Sibert III, "Theoretical studies of vibrationally excited polyatomic molecules using canonical Van Vleck perturbation theory," J. Chem. Phys. **88**(7), 4378–4390 (1988).
[18] V. Barone, "Anharmonic vibrational properties by a fully automated second-order perturbative approach," J. Chem. Phys. **122**(1), 014108 (2004).
[19] M. Piccardo, J. Bloino, and V. Barone, "Generalized vibrational perturbation theory for rotovibrational energies of linear, symmetric and asymmetric tops: Theory, approximations, and automated approaches to deal with medium-to-large molecular systems," Int. J. Quantum Chem. **115**(15), 948–982 (2015).
[20] R.B. Gerber, and M.A. Ratner, "A semiclassical self-consistent field (SC SCF) approximation for eigenvalues of coupled-vibration systems," Chem. Phys. Lett. **68**(1), 195–198 (1979).
[21] J.M. Bowman, "Self-consistent field energies and wavefunctions for coupled oscillators," J. Chem. Phys. **68**(2), 608–610 (1978).
[22] J.M. Bowman, "The self-consistent-field approach to polyatomic vibrations," Acc. Chem. Res. **19**(7), 202–208 (1986).
[23] G.M. Chaban, J.O. Jung, and R.B. Gerber, "Ab initio calculation of anharmonic vibrational states of polyatomic systems: Electronic structure combined with vibrational self-consistent field," J. Chem. Phys. **111**(5), 1823–1829 (1999).
[24] O. Christiansen, "Selected new developments in vibrational structure theory: potential construction and vibrational wave function calculations," Phys. Chem. Chem. Phys. **14**(19), 6672–6687 (2012).
[25] R.J. Whitehead, and N.C. Handy, "Variational calculation of vibration-rotation energy levels for triatomic molecules," J. Mol. Spectrosc. **55**(1), 356–373 (1975).
[26] G. Rauhut, "Configuration selection as a route towards efficient vibrational configuration interaction calculations," J. Chem. Phys. **127**(18), 184109 (2007).
[27] T. Carrington Jr., "Perspective: Computing (ro-)vibrational spectra of molecules with more than four atoms," J. Chem. Phys. **146**(12), 120902 (2017).
[28] A.G. Császár, C. Fábri, T. Szidarovszky, E. Mátyus, T. Furtenbacher, and G. Czakó, "The fourth age of quantum chemistry: molecules in motion," Phys. Chem. Chem. Phys. **14**(3), 1085–1106 (2012).
[29] P.M. Felker, and Z. Bačić, "Noncovalently bound molecular complexes beyond diatom–diatom systems: full-dimensional, fully coupled quantum calculations of rovibrational states," Phys. Chem. Chem. Phys. **24**(40), 24655–24676 (2022).
[30] G. Avila, and T. Carrington, "Nonproduct quadrature grids for solving the vibrational Schrödinger equation," J. Chem. Phys. **131**(17), 174103 (2009).

[31] D. Lauvergnat, and A. Nauts, "Torsional energy levels of nitric acid in reduced and full dimensionality with ElVibRot and Tnum," Phys. Chem. Chem. Phys. **12**(29), 8405–8412 (2010).
[32] S.N. Yurchenko, W. Thiel, and P. Jensen, "Theoretical ROVibrational Energies (TROVE): A robust numerical approach to the calculation of rovibrational energies for polyatomic molecules," J. Mol. Spectrosc. **245**(2), 126–140 (2007).
[33] G. Avila, and T. Carrington Jr., "Solving the Schroedinger equation using Smolyak interpolants," J. Chem. Phys. **139**(13), 134114 (2013).
[34] E. Castro, G. Avila, S. Manzhos, J. Agarwal, H.F. Schaefer, and T.C. Jr, "Applying a Smolyak collocation method to Cl2CO," Mol. Phys. **115**(15–16), 1775–1785 (2017).
[35] F.M. Hoffmann, "Infrared reflection-absorption spectroscopy of adsorbed molecules," Surf. Sci. Rep. **3**(2), 107–192 (1983).
[36] Y.J. Chabal, "Surface infrared spectroscopy," Surf. Sci. Rep. **8**(5), 211–357 (1988).
[37] K. Hadjiivanov, "Chapter Two - Identification and Characterization of Surface Hydroxyl Groups by Infrared Spectroscopy," in *Advances in Catalysis*, edited by F.C. Jentoft, (Academic Press, 2014), pp. 99–318.
[38] D.J. Cumming, C. Tumilson, S.F.R. Taylor, S. Chansai, A.V. Call, J. Jacquemin, C. Hardacre, and R.H. Elder, "Development of a diffuse reflectance infrared fourier transform spectroscopy (DRIFTS) cell for the in situ analysis of co-electrolysis in a solid oxide cell," Faraday Discuss. **182**, 97–111 (2015).
[39] A.M. McCullagh, R. Warringham, C.G.A. Morisse, L.F. Gilpin, C. Brennan, C.J. Mitchell, and D. Lennon, "A Comparison of Experimental Procedures for the Application of Infrared Spectroscopy to Probe the Surface Morphology of an Alumina-Supported Palladium Catalyst," Top. Catal. **64**(17), 1010–1020 (2021).
[40] W.C. Wilfong, C.S. Srikanth, and S.S.C. Chuang, "In Situ ATR and DRIFTS Studies of the Nature of Adsorbed CO2 on Tetraethylenepentamine Films," ACS Appl. Mater. Interfaces **6**(16), 13617–13626 (2014).
[41] M. Chan, T. Carrington, and S. Manzhos, "Anharmonic vibrations of the carboxyl group in acetic acid on TiO2: implications for adsorption mode assignment in dye-sensitized solar cells," Phys. Chem. Chem. Phys. **15**(25), 10028–10034 (2013).
[42] M.P. Andersson, and P. Uvdal, "First-Principle Calculations of the Experimental Vibrational Spectrum of a Surface Adsorbate: Anharmonic Resonance Coupling between Fundamental and Binary Modes," Phys. Rev. Lett. **90**(7), 076103 (2003).
[43] R. Franchy, M. Wuttig, and H. Ibach, "Anharmonic effects on adsorbate vibrations," Surf. Sci. **203**(3), 489–499 (1988).
[44] S.K. Chulkov, and D.M. Benoit, "A fragment method for systematic improvement of anharmonic adsorbate vibrational frequencies: Acetylene on Cu(001)," J. Chem. Phys. **139**(21), 214704 (2013).
[45] A. Wierzbicki, and J.M. Bowman, "Vibrational self-consistent field investigation of adsorbate vibrations: Coupling of atomic adsorbates to a rigid corrugated surface," Surface Science **191**(3), 518–528 (1987).
[46] G. Piccini, and J. Sauer, "Effect of Anharmonicity on Adsorption Thermodynamics," J. Chem. Theory Comput. **10**(6), 2479–2487 (2014).
[47] A. Erba, J. Maul, M. Ferrabone, P. Carbonnière, M. Rérat, and R. Dovesi, "Anharmonic Vibrational States of Solids from DFT Calculations. Part I: Description of the Potential Energy Surface," J. Chem. Theory Comput. **15**(6), 3755–3765 (2019).

[48] A. Erba, J. Maul, M. Ferrabone, R. Dovesi, M. Rérat, and P. Carbonnière, "Anharmonic Vibrational States of Solids from DFT Calculations. Part II: Implementation of the VSCF and VCI Methods," J. Chem. Theory Comput. **15**(6), 3766–3777 (2019).
[49] S. Manzhos, M. Ihara, and T. Carrington, "Using Collocation to Solve the Schrödinger Equation," J. Chem. Theory Comput. **19**(6), 1641–1656 (2023).
[50] T. Carrington, "Using collocation to study the vibrational dynamics of molecules," Spectrochimica Acta Part A: Molecular and Biomolecular Spectroscopy **248**, 119158 (2021).
[51] S. Manzhos, and T. Carrington, "Using an internal coordinate Gaussian basis and a space-fixed Cartesian coordinate kinetic energy operator to compute a vibrational spectrum with rectangular collocation," J. Chem. Phys. **145**(22), 224110 (2016).
[52] W. Yang, and A.C. Peet, "The collocation method for bound solutions of the Schrödinger equation," Chem. Phys. Lett. **153**(1), 98–104 (1988).
[53] S. Manzhos, and T. Carrington, "Advances in rectangular collocation for solution of the Schrödinger equation: from obviating integrals to machine learning," in *Density Functionals for Many-Particle Systems*, (WORLD SCIENTIFIC, 2022), pp. 319–344.
[54] S.F. Boys, "Some bilinear convergence characteristics of the solutions of dissymmetric secular equations," Proc. Royal Soc. London A **309**(1497), 195–208 (1969).
[55] J. Simmons, and T. Carrington Jr., "Computing vibrational spectra using a new collocation method with a pruned basis and more points than basis functions: Avoiding quadrature," J. Chem. Phys. **158**(14), 144115 (2023).
[56] S. Garashchuk, and J.C. Light, "Quasirandom distributed Gaussian bases for bound problems," J. Chem. Phys. **114**(9), 3929–3939 (2001).
[57] Z. Bačić, and J.C. Light, "Highly excited vibrational levels of '"floppy"' triatomic molecules: A discrete variable representation—Distributed Gaussian basis approach," J. Chem. Phys. **85**(8), 4594–4604 (1986).
[58] I.P. Hamilton, and J.C. Light, "On distributed Gaussian bases for simple model multidimensional vibrational problems," J. Chem. Phys. **84**(1), 306–317 (1986).
[59] S. Manzhos, T. Carrington, L. Laverdure, and N. Mosey, "Computing the Anharmonic Vibrational Spectrum of UF6 in 15 Dimensions with an Optimized Basis Set and Rectangular Collocation," J. Phys. Chem. A **119**(36), 9557–9567 (2015).
[60] S. Manzhos, M. Chan, and T. Carrington, "Communication: Favorable dimensionality scaling of rectangular collocation with adaptable basis functions up to 7 dimensions," J. Chem. Phys. **139**(5), 051101 (2013).
[61] M. Chan, S. Manzhos, T. Carrington, and K. Yamashita, "Parameterized Bases for Calculating Vibrational Spectra Directly from ab Initio Data Using Rectangular Collocation," J. Chem. Theory Comput. **8**(6), 2053–2061 (2012).
[62] S. Manzhos, X. Wang, and T. Carrington, "A multimode-like scheme for selecting the centers of Gaussian basis functions when computing vibrational spectra," Chem. Phys. **509**, 139–144 (2018).
[63] J. Ku, A. Kamath, T. Carrington, and S. Manzhos, "Machine Learning Optimization of the Collocation Point Set for Solving the Kohn–Sham Equation," J. Phys. Chem. A **123**(49), 10631–10642 (2019).

[64] S. Manzhos, and T. Carrington, "Using rectangular collocation with finite difference derivatives to solve electronic Schrödinger equation," J. Chem. Phys. **149**(20), 204105 (2018).
[65] A. Kamath, and S. Manzhos, "Inverse Multiquadratic Functions as the Basis for the Rectangular Collocation Method to Solve the Vibrational Schrödinger Equation," Mathematics **6**(11), 253 (2018).
[66] S. Manzhos, K. Yamashita, and T. Carrington, "On the advantages of a rectangular matrix collocation equation for computing vibrational spectra from small basis sets," Chem. Phys. Lett. **511**(4), 434–439 (2011).
[67] B. Poirier, and J.C. Light, "Efficient distributed Gaussian basis for rovibrational spectroscopy calculations," J. Chem. Phys. **113**(1), 211–217 (2000).
[68] S.W. Flynn, and V.A. Mandelshtam, "Molecular Spectra Calculations Using an Optimized Quasi-Regular Gaussian Basis and the Collocation Method," J. Chem. Theory Comput. **17**(11), 7169–7177 (2021).
[69] H. Weyl, "Über die Gleichverteilung von Zahlen mod. Eins," Math. Ann. **77**(3), 313–352 (1916).
[70] I.M. Sobol', "On the distribution of points in a cube and the approximate evaluation of integrals," USSR Comput. Math. Math. Phys. **7**(4), 86–112 (1967).
[71] T. Carrington, and R.G. Owens, "Reducing least-squares and collocation error using a condition-number bound on a generalized Lebesgue constant," ChemRxiv **2026**(0515), (n.d.).
[72] D. Lauvergnat, and A. Nauts, "Exact numerical computation of a kinetic energy operator in curvilinear coordinates," J. Chem. Phys. **116**(19), 8560–8570 (2002).
[73] P. Hohenberg, and W. Kohn, "Inhomogeneous Electron Gas," Phys. Rev. **136**(3B), B864–B871 (1964).
[74] W. Kohn, and L.J. Sham, "Self-Consistent Equations Including Exchange and Correlation Effects," Phys. Rev. **140**(4A), A1133–A1138 (1965).
[75] A.D. Becke, "Density-functional thermochemistry. III. The role of exact exchange," J. Chem. Phys. **98**(7), 5648–5652 (1993).
[76] Q. Sun, T.C. Berkelbach, N.S. Blunt, G.H. Booth, S. Guo, Z. Li, J. Liu, J.D. McClain, E.R. Sayfutyarova, S. Sharma, S. Wouters, and G.K.-L. Chan, "PySCF: the Python-based simulations of chemistry framework," WIREs Comp. Mol. Sci. **8**(1), e1340 (2018).
[77] Q. Sun, X. Zhang, S. Banerjee, P. Bao, M. Barbry, N.S. Blunt, N.A. Bogdanov, G.H. Booth, J. Chen, Z.-H. Cui, J.J. Eriksen, Y. Gao, S. Guo, J. Hermann, M.R. Hermes, K. Koh, P. Koval, S. Lehtola, Z. Li, J. Liu, N. Mardirossian, J.D. McClain, M. Motta, B. Mussard, H.Q. Pham, A. Pulkin, W. Purwanto, P.J. Robinson, E. Ronca, E.R. Sayfutyarova, M. Scheurer, H.F. Schurkus, J.E.T. Smith, C. Sun, S.-N. Sun, S. Upadhyay, L.K. Wagner, X. Wang, A. White, J.D. Whitfield, M.J. Williamson, S. Wouters, J. Yang, J.M. Yu, T. Zhu, T.C. Berkelbach, S. Sharma, A.Yu. Sokolov, and G.K.-L. Chan, "Recent developments in the PySCF program package," J. Chem. Phys. **153**(2), 024109 (2020).
[78] G.D. Purvis III, and R.J. Bartlett, "A full coupled-cluster singles and doubles model: The inclusion of disconnected triples," J. Chem. Phys. **76**(4), 1910–1918 (1982).
[79] K. Raghavachari, G.W. Trucks, J.A. Pople, and M. Head-Gordon, "A fifth-order perturbation comparison of electron correlation theories," Chem. Phys. Lett. **157**(6), 479–483 (1989).

[80]"Claude Opus 5.5 [large language model]," (n.d.).
[81] G.E. Rossi, J.M. Winfield, C.J. Mitchell, W. van der Borden, K. van der Velde, R.H. Carr, and D. Lennon, "Phosgene formation via carbon monoxide and dichlorine reaction over an activated carbon catalyst: Reaction testing arrangements," Appl. Catal. A **594**, 117467 (2020).
[82] R. Burcl, S. Carter, and N.C. Handy, "On the representation of potential energy surfaces of polyatomic molecules in normal coordinates: II. Parameterisation of the force field," Chemical Physics Letters **373**(3), 357–365 (2003).
[83] K. Yagi, K. Hirao, T. Taketsugu, M.W. Schmidt, and M.S. Gordon, "Ab initio vibrational state calculations with a quartic force field: Applications to H2CO, C2H4, CH3OH, CH3CCH, and C6H6," J. Chem. Phys. **121**(3), 1383–1389 (2004).
[84] Z.H. Kafafi, R.H. Hauge, W.E. Billups, and J.L. Margrave, "Carbon dioxide activation by lithium metal. 1. Infrared spectra of lithium carbon dioxide (Li+CO2-), lithium oxalate (Li+C2O4-), and lithium carbon dioxide (Li22+CO22-) in inert-gas matrices," J. Am. Chem. Soc. **105**(12), 3886–3893 (1983).